\documentclass[graybox]{svmult}

\usepackage{type1cm}        
\usepackage{makeidx}         
\usepackage{graphicx}        
\usepackage{multicol}        
\usepackage[bottom]{footmisc}

\usepackage{epsfig}
\usepackage{epstopdf}
\usepackage{newtxtext}       %
\usepackage{newtxmath}       

\makeindex             

\usepackage{bm}
\usepackage{hyperref}
\allowdisplaybreaks[4]
\graphicspath{{./images/} {./results/} {./bubble/} {./RMI/}}
\usepackage[justification=centering]{caption}

\theoremstyle{plain}

\begin{document}

\title*{An energy-splitting high order numerical method for multi-material flows}
\author{Xin Lei and Jiequan Li}
\institute{Xin Lei \at School of Science, China University of Geosciences, \email{xinlei@cugb.edu.cn}
\and Corresponding to:  Jiequan Li \at  Laboratory of Computational Physics, Institute of Applied Physics and Computational Mathematics, Beijing; and Center for Applied Physics and Technology, Peking University, \email{li\_jiequan@iapcm.ac.cn}}
%
%
\maketitle


\abstract{This chapter deals with multi-material flow problems by a kind of  effective numerical methods, based on a series of reduced forms of the Baer-Nunziato (BN) model.
Numerical simulations often face a host of difficult challenges, typically including the volume fraction positivity and stability  of multi-material shocks.
To cope with these challenges, we propose a new non-oscillatory {\em energy-splitting} Godunov-type scheme for computing multi-fluid flows in the Eulerian framework.
A novel reduced version of the BN model is introduced  as the basis for the energy-splitting scheme.
In comparison with existing two-material compressible flow models obtained by reducing the BN model in the literature, it is shown that our new reduced model can simulate the kinetic energy exchange around material interfaces very effectively.
Then  a second-order accurate extension of the energy-splitting  Godunov-type scheme is made  using the generalized Riemann problem (GRP) solver.
Numerical experiments are carried out for the shock-interface interaction, shock-bubble interaction and the Richtmyer-Meshkov instability problems, which  demonstrate the excellent performance of this type of schemes.
}

\section{Introduction}\label{sec:intr}

There are three typical frameworks for numerically simulating multi-material flows: the Lagrangian framework, the Eulerian framework and the ALE (Arbitrary Lagrangian-Eulerian) framework.
The ALE framework involves a compromise between the first two frameworks, where the Lagrangian part provides the treatment of material interfaces.

The Lagrangian formulation of fluid dynamics uses a coordinate system moving with the velocity field of the fluid \cite{francois_comparative_2013,maire_cell-centered_2007,cheng_positivity-preserving_2014}.
For multi-material flows, Lagrangian methods can track material interfaces in real time and keep the interfaces sharp. It is preferable in the compressible multi-material flows due to its sharp capturing ability of material interfaces. 
However, it should be noted that in the Lagrangian framework, excessive distortion and tangle of the mesh may lead to the degradation of the numerical solution \cite{chang_compatible_2012}.
Moreover, the inflow or outflow of the multi-material flow corresponds to the increase or decrease of grid nodes, which causes difficulties in implementation in practice. Therefore, the pure Lagrangian method is not suitable for general multi-fluid problems.

Eulerian methods are essentially shock-capturing methods over a fixed mesh, which simulate inviscid multi-material flows using the Euler equations.
This type of methods belongs to the family of  diffusion interface methods, which can produce diffusive solutions and  allow different materials to artificially mix in a small region.
There are usually two approaches to construct models appropriate for Eulerian methods: one approach is based on some augmented form of the Euler equations for the multi-material mixture, the other approach is the multiphase approach \cite{michael_hybrid_2016}.
The augmented Euler approach identifies each material in the mixing region through one or more additional evolution equations.
This kind of model in the one-dimensional case usually consists of four equations, also known as the four-equation multi-component flow model.
Widely used augmented Euler formulations include additional evolution equations describing the mass fraction or volume fraction of a material, or the ratio of specific heats $\gamma$ for the mixture.
The corresponding models are the mass fraction model \cite{banks_high-resolution_2007,larrouturou_how_1991,quirk_dynamics_1996} (including the reactive flows with mass fractions of burnt/unburnt materials \cite{ben-artzi_computation_1990,ben-artzi_generalized_1989}), the volume fraction model \cite{shyue_efficient_1998} and the $\gamma$-model \cite{abgrall_how_1994,karni_multicomponent_1994,shyue_efficient_1998}.
In the mixing region with more than one material across material interfaces, an equation of state for the mixture is required for such a task.
The Eulerian numerical method based on the conservative augmented Euler system is very sensitive to the mixing rules to close the systems, and incompatibility of the mixing rules may cause spurious pressure oscillations near material interfaces.
Therefore, some single-fluid algorithms and quasi-conservative or non-conservative approaches were suggested to ensure the correct numerical fluid mixing rules at interfaces.
Some representative non-conservative approaches to prevent spurious oscillations include the correction of the energy inside computational cells \cite{jenny_correction_1997, banks_high-resolution_2007} and a pressure evolution model to compute the pressure directly.


The multiphase approach considers the full or reduced forms of the Baer-Nunziato (BN) model \cite{baer_two-phase_1986}. 
In the complete BN model, additional governing equations of different materials are necessary, and this model is a  seven-equation model.
Reduced forms of the BN model capture more physical information than the augmented Euler approach does.
However, such models are more complex and require greater computational costs.
We will continue to consider such approaches in the next section.


A comparative study of various multi-material Lagrangian and Eulerian methods was made in \cite{francois_comparative_2013}.
The shortcomings of Eulerian and Lagrangian methods drive the development of other strategies that take the advantages of each method and attempt to avoid these drawbacks.

Accurate tracking of sharp material interface is important in the simulation of multi-material flows.
However, it is difficult to maintain sharp material interfaces in Eulerian meshes.
Even so, there are various ways to preserve the sharpness of material interfaces for Eulerian simulations, such as front-tracking methods, moving boundary tracking (MBT) methods  \cite{falcovitz_two-dimensional_1997,shyue_moving-boundary_2008, Du-Li-2020}, level-set methods \cite{karni_multicomponent_1994,karni_hybrid_1996,mulder_computing_1992} and 
interface reconstruction methods.
Front-tracking methods take discontinuities (including material interfaces) as moving fronts and use an additional Lagrangian grid on the interfaces.
 For example, a front-tracking method was proposed,  involving a standard Eulerian predictor step and a corrector step applied only to grid cells near the interface \cite{cocchi_riemann_1997}.  Recently the front-tracking method was improved analytically to be fourth order accurate via mapping and adjusting regular semialgebraic sets \cite{Zhang-MARS}. Thus, the material regions are broadly modeled by point sets with piecewise smooth boundaries and passively advected by flow maps determined from the velocity. 
The MBT methods treat material interfaces in compressible multi-material flows as moving impermeable boundary surfaces embedded in a regular Eulerian mesh.
The motion of boundary surfaces is treated in Lagrangian coordinates, and the motion of immiscible dynamically interacting materials away from boundary surfaces are represented in terms of  Eulerian coordinates.
In two space dimensions, a classical MBT method with the application of a cell merging technique and the dimensional splitting GRP scheme was developed in \cite{falcovitz_two-dimensional_1997}. A basic one-dimensional (1-D) form of this MBT method was proposed in \cite{ben-artzi_application_1986}.
In the above two methods, the Eulerian portion avoids mesh deformation, but it is difficult to track the newly created and imperfectly located material interfaces in the flow field.
To locate newly created interfaces, the level-set method is adopted in the simulation of multi-fluid flows, using the zero level set of the discrete signed distance function to identify different materials.
The exact mixing concentration in the cells across the material interfaces is  unavailable by this method,  which will produce the numerical performance of non-physical  pressure oscillations around interfaces \cite{karni_hybrid_1996,karni_multicomponent_1994}. The ghost fluid method (GFM) was a representative of the  modified level set method with excellent numerical performance  \cite{fedkiw_non-oscillatory_1999}  and an adaptive mesh refinement extension of GFM can be found in \cite{nourgaliev_adaptive_2006, liu_ghost_2003}.
The most popular interface reconstruction method is the volume-of-fluid (VOF) method \cite{colella_multifluid_1989,miller_high-order_1996,ton_improved_1996}, which reconstruct the interface location at each time step from the volume fractions of different materials in each cell. In \cite{Xiao-THINC-2011} the hyperbolic tangent function  was introduced to   greatly improve the quality of material interfaces  in terms of sharpness.  
The moment-of-fluid (MOF) method, another interface reconstruction method,  was proposed \cite{ahn_moment--fluid_2009,dyadechko_moment--fluid_2005} by using the volume fraction and centroid for a more accurate representation of the material configuration, interfaces and volume advection.
Complex changes of the interface topology are handled easily using level set methods and interface reconstruction methods.


In contrast with Lagrangian and Eulerian frameworks, 
the idea of the ALE methods \cite{kamm_pressure_2009,hirt_arbitrary_1974,galera_two-dimensional_2010,ben-artzi_application_1986} is that the mesh motion can be arbitrarily selected, and the freedom of mesh motion provides additional flexibility and accuracy of the simulation.
Usually, ALE methods can be implemented in two manners.
One is the direct ALE  involving an unsplit moving mesh discretization of the governing equations, and the other is the indirect ALE.
In the practical computation, the key point of the indirect ALE methods is accurately remapping all fluid variables from the Lagrangian solution to the rezoned mesh.
Under the ALE frameworks, there might be mixed cells that contain more than one materials.
Each material has its own density and may have its own internal energy and pressure.
To close the system of the governing equations and produce the pressure and other variables of each material, pressure equilibrium or pressure relaxation techniques are usually used.
In \cite{shashkov_closure_2008}, there is another class of closure methods for constructing closed models based on modeling the dynamics of the pure subcells within a mixed cell.
A relaxation-projection method by Lagrangian plus remapping the flow quantities to the Eulerian mesh was designed in  \cite{saurel_relaxation-projection_2007}.
However, for the ALE method, the drawbacks of Lagrangian and Eulerian methods can not be completely avoided. Since Eulerian mesh is preferred to be used at material interfaces to prevent tangled meshes, the mixing rules on the mixing region still need to be set precisely.

In addition, most of the above Eulerian and ALE methods are not completely conservative, such as the correction algorithm of energy, the GFM/VOF/MOF method and the relaxation-projection method, and then errors may generate in the simulation of multi-material shocks by these methods.
According to the descriptions above, in the computation of compressible multi-material flows and multiphase mixtures, a lot of problems may arise, awaiting for  resolution,  as follows. 
\begin{itemize}
\item The presence of spurious oscillations in velocity and pressure across material interfaces. 
\item The convergence issue involving shocks: The simulations of shocks in mixture regions is questionable since the convergence of numerical  solutions of a non-conservative model is extremely difficult.
\item Positivity of the volume fraction (or the phasic internal energy), especially in the case that material interfaces interact with shocks or strong expansion waves, or when moving interfaces appear in almost pure liquids and solids.
\item Extensibility to high-order accuracy and multi-dimensionality.
\end{itemize}

In fact, the seven-equation multiphase flow models in the multiphase approach are usually not prone to the above problems.
Nevertheless, since the seven-equation models contain very complex and diverse wave structures and are sensitive to the relaxation process, the numerical simulation of this kind of model is complicated.
For the most simplified case, the four-equation model are widely used \cite{michael_hybrid_2016}. 
However, the four-equation models often confront the interface oscillation problem for conservative formulations or shock simulation problem for non-conservative formulations.
The best approach is to strike a balance between complexity and veracity of the models. 


Our energy-splitting method is a kind of the multi-material approach based on the reduced model and takes account of the  exchange of kinetic energy  and internal energy inside a mixing Eulerian cell \cite{lei_non-oscillatory_2018}.  The extension to second order accuracy uses the the generalized Riemann problem (GRP) solver \cite{BF_1984, ben-artzi_generalized_2003,ben-artzi_direct_2006,Li-2}.  The method is purely built over an Eulerian mesh by inspecting the kinetic energy exchange between different materials   inside an Eulerian cell. 
The organization of this paper basically  consists of five sections. Besides the introduction here, Section \ref{sec:stiff-relax} describes the connection of different multi-material models that  are all derived from the BN model.
In Section 3 we summarize the numerical methods for these reduced models and in Section 4, a number of numerical results are displayed to demonstrate the performance of the our proposed method. The conclusion is drawn in Section 5. Finally we put in Appendix the GRP solver for completeness of presentation.

\section{Two-material compressible flow models}\label{sec:stiff-relax}

\subsection{Seven-equation model}

First of all, we consider a class of two-velocity two-pressure models for two materials in 1-D space, which is often called the "seven-equation" model. It was first proposed by Baer and Nunziato in \cite{baer_two-phase_1986} to describe the deflagration-to-detonation transition in porous granular explosives mixed with gaseous products of combustion. Subsequently, this model was extended to simulate gas-liquid flows in \cite{saurel_multiphase_1999}.
In this class of models, the governing equations for two materials $a,b$ and their mixture read:
\begin{subequations}\label{cons-eqns}
\begin{align}
&\frac{\partial \rho \phi_{a}}{\partial t}+\operatorname{div}\left(\rho \phi_{a} \bm{u}\right)=0,\\
&\frac{\partial \rho \phi_{a} \bm{u}_{a}}{\partial t}+\operatorname{div}\left(\rho \phi_{a} \bm{u}_{a} \otimes \bm{u}_{a}\right)+\nabla (z_{a} p_{a})=p_\mathrm{I} \nabla z_{a}+\lambda\left(\bm{u}_{b}-\bm{u}_{a}\right),\\
&\frac{\partial \rho \phi_{a} E_{a}}{\partial t}+\operatorname{div}\left[\left(\rho \phi_{a} E_{a}+z_{a} p_{a}\right) \bm{u}_{a}\right]=-p_\mathrm{I} \frac{\partial z_{a}}{\partial t}+\lambda \bm{u}_\mathrm{I} \cdot\left(\bm{u}_{b}-\bm{u}_{a}\right),\\
&\frac{\partial \rho \phi_{b}}{\partial t}+\operatorname{div}\left(\rho \phi_{b} \bm{u}\right)=0,\\
&\frac{\partial \rho \phi_{b} \bm{u}_{b}}{\partial t}+\operatorname{div}\left(\rho \phi_{b} \bm{u}_{b} \otimes \bm{u}_{b}\right)+\nabla (z_{b} p_{b})=p_\mathrm{I} \nabla z_{b}-\lambda\left(\bm{u}_{b}-\bm{u}_{a}\right),\\
&\frac{\partial \rho \phi_{b} E_{b}}{\partial t}+\operatorname{div}\left[\left(\rho \phi_{b} E_{b}+z_{b} p_{b}\right) \bm{u}_{b}\right]= p_\mathrm{I} \frac{\partial z_{a}}{\partial t}-\lambda \bm{u}_\mathrm{I} \cdot\left(\bm{u}_{b}-\bm{u}_{a}\right),\\
&\frac{\partial z_{a}}{\partial t}+\bm{u} \cdot \nabla z_{a}=\mu (p_a-p_b).
\end{align}
\end{subequations}
For the material $k$, $k=a$ or $b$, the density, velocity, pressure, specific internal energy and specific total energy are denoted as
$\rho_k=\rho \phi_k\big / z_k$, $\bm{u}_k$, $p_k$, $e_k$ and $E_k=e_k+\bm{u}_k^2\big/2$, respectively.
The volume fractions $z_k$ and mass fraction $\phi_k$ satisfy the saturation constraint
\begin{equation}\label{eq:sum-1}
z_a+z_b =1,\quad
\phi_a + \phi_b =1.
\end{equation}
In accord with  conservation properties, the density $\rho$, velocity $\bm{u}$ and specific total energy $E$ for the mixture satisfy
\begin{equation}\label{eq:cons-prop}
\rho = z_a \rho_a + z_b \rho_b,\quad
\bm{u} = \phi_a\bm{u}_a+\phi_b\bm{u}_b,\quad
E = \phi_a E_a+\phi_b E_b,
\end{equation}
and the pressure $p$ for the mixture complies with the Dalton's law of partial pressures
\begin{equation}\label{eq:Dalton}
p = z_a p_a + z_b p_b.
\end{equation}
The pressure $p_\mathrm{I}$ and velocity $\bm{u}_\mathrm{I}$ represent averaged
values of the interfacial pressure and velocity over the control volume of the mixture.
Taking the form of the interfacial variables in \cite{saurel_multiphase_1999} as an example, we set $p_\mathrm{I}=p$ and $\bm{u}_\mathrm{I}=\bm{u}$.
The relaxation coefficient of velocity and pressure are $\lambda$ and $\mu$, respectively.
These two coefficients reflect the rate at which velocity and pressure converge to equilibrium.


To simplify the complexity of the seven-equation model, we next consider some reduced models of \eqref{cons-eqns} in the limit of stiff velocity relaxation ($\lambda\rightarrow \infty$) and stiff pressure relaxation ($\mu\rightarrow \infty$).
In many physical situations, estimates reveal that the length scales for
velocity and pressure equilibration are very small, thus the assumption of stiff velocity relaxation is reasonable.
These two limiting cases correspond to instantaneous velocity equilibrium and pressure equilibrium, respectively.
Let us start with the limiting case of stiff velocity relaxation.


\subsection{Six-equation model}

In this section, only the stiff velocity relaxation is considered.
We eliminate the velocity relaxation term $\lambda (\bm{u}_b-\bm{u}_a)$ in (\ref{cons-eqns}b) and (\ref{cons-eqns}c), and then get
\begin{equation}\label{eq:six-1}
\begin{aligned}
 & \frac{\partial (\rho \phi_{a} E_{a})}{\partial t}+\operatorname{div}\left[\left(\rho \phi_{a} E_{a}+z_{a} p_{a}\right) \bm{u}_{a}\right]\\
= \ &-p \frac{\partial z_{a}}{\partial t} + \bm{u} \cdot\left(\frac{\partial \rho \phi_{a} \bm{u}_{a}}{\partial t}+\operatorname{div}\left(\rho \phi_{a} \bm{u}_{a} \otimes \bm{u}_{a}\right)+\nabla (z_{a} p_{a})-p \nabla z_{a}\right).
\end{aligned}
\end{equation}
Let us specially label with superscript ``$*$'' the variables in the seven-equation system \eqref{cons-eqns} with stiff velocity relaxation. Then instantaneous velocity equilibrium means $\bm{u}_a^*=\bm{u}_b^*=\bm{u}$.
If (\ref{cons-eqns}g) is substituted into the above equation, \eqref{eq:six-1} becomes
\begin{equation}\label{eq:5-inter-eneg}
\frac{\partial (\rho \phi_{a} e_{a}^*)}{\partial t}+\operatorname{div}\left(\rho \phi_{a} e_{a}^* \bm{u}\right)+z_{a} p_{a}^* \operatorname{div} \bm{u} = -p\left(\frac{\partial z_{a}}{\partial t}+\bm{u} \cdot \nabla z_{a}\right) = -\mu p(p_a^*-p_b^*).
\end{equation}
The symmetric internal energy equation holds for the material $b$ too,
\begin{equation}\label{eq:5-inter-eneg-b}
\frac{\partial (\rho \phi_{b} e_{b}^*)}{\partial t}+\operatorname{div}\left(\rho \phi_{b} e_{b}^* \bm{u}\right)+z_{b} p_{b}^* \operatorname{div} \bm{u} = \mu p(p_a^*-p_b^*).
\end{equation}
In addition, the sum of (\ref{cons-eqns}b) and (\ref{cons-eqns}d) is the momentum equation for the mixture,
\begin{equation}\label{eq:5-momen}
\frac{\partial \rho \bm{u}}{\partial t}+\operatorname{div}(\rho \bm{u} \otimes \bm{u})+\nabla p=0.
\end{equation}
The system composed of (\ref{cons-eqns}a), (\ref{cons-eqns}d), \eqref{eq:5-inter-eneg}, \eqref{eq:5-inter-eneg-b}, \eqref{eq:5-momen} and (\ref{cons-eqns}g) actually is the six-equation single-velocity model for two materials  \cite{richard_saurel_simple_2009,kapila_two-phase_2001}.

In order to make the energy equations for a single material in the model, corresponding to \eqref{eq:5-inter-eneg-b} and formally consistent with the total energy equation of the mixture, 
another six-equation model was proposed by replacing the internal energy equations with the total energy equations for two materials \cite{pelanti_mixture-energy-consistent_2014,kapila_two-phase_2001}.
Specifically, the stiff velocity relaxation term is calculated by the following steps.
According to (\ref{cons-eqns}a) and (\ref{cons-eqns}b), 
\begin{equation}\label{eq:6-another-a}
\rho \phi_{a} \frac{\operatorname{D} \bm{u}_a}{\operatorname{D} t}+\nabla (z_{a} p_{a})=p_I \nabla z_{a}+\lambda\left(\bm{u}_{b}-\bm{u}_{a}\right),\quad
\frac{\operatorname{D}(\bullet)}{\operatorname{D} t} =\frac{\partial(\bullet)}{\partial t} +\bm{u} \cdot \nabla (\bullet)
\end{equation}
holds for the material $a$, and the symmetric equation holds for the material $b$,
\begin{equation}\label{eq:6-another-b}
\rho \phi_{b} \frac{\operatorname{D} \bm{u}_b}{\operatorname{D} t}+\nabla (z_{b} p_{b})=p_I \nabla z_{b}-\lambda\left(\bm{u}_{b}-\bm{u}_{a}\right),
\end{equation}
where $\operatorname{D}/\operatorname{D} t$ is the material derivative. 
Then the combination of  \eqref{eq:6-another-a} and   \eqref{eq:6-another-b} derives the stiff velocity relaxation term
\begin{equation}\label{eq:stiff-velocity}
\lambda\left(\bm{u}_{b}^* -\bm{u}_{a}^*\right)=\rho \phi_{a}\phi_{b}  \frac{\operatorname{D} \left(\bm{u}_a^*-\bm{u}_b^*\right)}{\operatorname{D} t} + \Sigma^* - p\nabla z_a = \Sigma^* - p\nabla z_a
\end{equation}
for $\lambda\rightarrow \infty$, where $\Sigma=\phi_b\nabla (z_a p_a)-\phi_a\nabla (z_b p_b)$.
Hence, the total energy equation (\ref{cons-eqns}c) for the material $a$ is written as
\begin{equation}\label{eq:6-eq-internal-eneg}
\frac{\partial \rho \phi_{a} E_{a}^*}{\partial t}+\operatorname{div}\left[\left(\rho \phi_{a} E_{a}^*+z_{a} p_{a}^*\right) \bm{u}\right]=-\mu p(p_a^*-p_b^*) + \bm{u}\cdot\Sigma^*.
\end{equation}
To summarize, the total-energy form of the six-equation model is written as
\begin{subequations}\label{eq:6-eq}
\begin{align}
&\frac{\partial \rho \phi_{a}}{\partial t}+\operatorname{div}\left(\rho \phi_{a} \bm{u}\right)=0,\\
&\frac{\partial \rho \phi_{b}}{\partial t}+\operatorname{div}\left(\rho \phi_{b} \bm{u}\right)=0,\\
&\frac{\partial \rho \bm{u}}{\partial t}+\operatorname{div}(\rho \bm{u} \otimes \bm{u})+\nabla p=0,\\
&\frac{\partial \rho \phi_{a} E_{a}}{\partial t}+\operatorname{div}\left[\left(\rho \phi_{a} E_{a} +z_{a} p_{a}\right) \bm{u}\right] - \bm{u}\cdot\Sigma = - \mu p(p_a-p_b),\\
&\frac{\partial \rho \phi_{b} E_{b}}{\partial t}+\operatorname{div}\left[\left(\rho \phi_{b} E_{b} +z_{b} p_{b}\right) \bm{u}\right] + \bm{u}\cdot\Sigma = \mu p(p_a-p_b),\\
&\frac{\partial z_{a}}{\partial t}+\bm{u} \cdot \nabla z_{a}=\mu (p_a-p_b),
\end{align}
\end{subequations}
and in compact form as
\begin{equation}\label{eq:6-eq-compact}
\frac{\partial \mathcal{U}}{\partial t}+\operatorname{div}\mathcal{F}(\mathcal{U})+\bm{u}\cdot \mathcal{S}(\mathcal{U}) = \mu \mathcal{R}(\mathcal{U}),
\end{equation}
with the averaged sound speed $c$ for the mixture  defined by
\begin{equation}
c=\sqrt{\sum_{k=a,b}\phi_k c_k^2}.
\end{equation}
In this six-equation model, the stiff velocity relaxation term $\bm{u}\cdot\Sigma^*=\lambda \bm{u}\cdot(\bm{u}_b^*-\bm{u}_a^*)$ reflects the kinetic energy exchange between two materials induced by the instantaneous velocity equilibrium.
However, we find that $\Sigma$ is formally independent of the velocity field $\bm{u}$. Hence, it is arduous to completely capture the kinetic energy exchange by only appropriate numerical discretization of $\bm{u}\cdot\Sigma$ in \eqref{eq:6-eq-internal-eneg}.

In addition, the six-equation model is a pressure non-equilibrium model which is sensitive to the pressure relaxation process.
We believe that the  consideration of the instantaneous pressure equilibrium will have a greater impact on the model. 
Next, we introduce the classical five-equation model with stiff pressure relaxation.
To better simulate the kinetic energy exchange, another form of the total energy equation for the material $a$ with stiff pressure relaxation is presented in Section \ref{sec:new-reduced}.


\subruninhead{The $\gamma$-law.}
For the polytropic gases, the equations of state (EOS) for the material $k$ and the mixture are
\begin{equation}\label{ideal_EOS}
p_k = (\gamma_k -1)\rho_k e_k,\quad
p = (\gamma -1)\rho e,
\end{equation}
where $\gamma_k=C_{p,k}/C_{v,k}$ is the ratio of the specific heat capacity at constant pressure and volume, respectively, of the material $k$,
and the specific internal energy $e$ for the mixture under the velocity equilibration satisfies
\begin{equation}\label{eq:cons-prop2}
e = \phi_a e_a+\phi_b e_b.
\end{equation}
According to \eqref{eq:sum-1}, \eqref{eq:Dalton} and \eqref{eq:cons-prop2}, the ratio of specific heats for mixture is obtained as
\begin{equation}\label{gamma}
\gamma = \gamma(\phi_a, e_a, e_b) = \frac{\sum_{k=a,b} \phi_k e_k \gamma_k}{\sum_{k=a,b} \phi_k e_k}.
\end{equation}

\subsection{Five-equation reduced model}

Based on an asymptotic analysis in the limit of stiff velocity and pressure relaxation, a five-equation reduced model   was proposed in \cite{murrone_five_2005,kapila_two-phase_2001}. 
The following is the derivation of this model.

Let us start with the six-equation model under instantaneous velocity equilibrium.
The equation (\ref{eq:6-eq}a) is equivalent to
\begin{equation}
\frac{\partial z_{a} \rho_a}{\partial t}+\operatorname{div}\left(z_{a} \rho_{a} \bm{u}\right)=0,
\end{equation}
and thus the density of material $a$ obeys
\begin{equation}\label{eq:D-Dt1}
z_a \frac{\operatorname{D} \rho_a}{\operatorname{D} t}+z_{a} \rho_a \operatorname{div}\bm{u}=-\rho_a \frac{\operatorname{D} z_{a}}{\operatorname{D} t}.
\end{equation}
The equation \eqref{eq:5-inter-eneg} can be written as
\begin{equation}\label{eq:D-Dt2}
z_a \rho_a \frac{\operatorname{D} e_{a}^*}{\operatorname{D} t}+z_{a} p_{a}^*\operatorname{div} \bm{u} = -p \frac{\operatorname{D} z_{a}}{\operatorname{D} t}.
\end{equation}
According to the thermodynamic relations
\begin{equation}\label{eq:second-law}
\mathrm{d}e_k=T_k \mathrm{d}S_k+\frac{p_k}{\rho_k^2}\mathrm{d}\rho_k,\quad
\mathrm{d}p_k=c_k^2 \mathrm{d}\rho_k+\left(\frac{\partial p_k}{\partial S_k}\right)_{\rho_k}\mathrm{d}S_k,
\end{equation}
where $c_k=\sqrt{\left(\partial p_k/\partial \rho_k\right)_{S_k}}$ is the sound speed of the material $k$, 
\eqref{eq:D-Dt1} and \eqref{eq:D-Dt2} imply the equation for the entropy of the material $a$,
\begin{equation}\label{eq:D-Dt3}
\frac{\operatorname{D} S_{a}^*}{\operatorname{D} t} = 0
\quad \text{and} \quad
\frac{\operatorname{D} p_{a}^*}{\operatorname{D} t} = (c_a^*)^2 \frac{\operatorname{D} \rho_{a}}{\operatorname{D} t}.
\end{equation}
Substituting them into \eqref{eq:D-Dt2}, we obtain
\begin{equation}\label{eq:5-eq-a}
\frac{\operatorname{D} p_b^*}{\operatorname{D} t}+\rho_b (c_b^*)^2 \operatorname{div}\bm{u}=-\frac{\rho_b (c_b^*)^2}{z_b} \frac{\operatorname{D} z_{b}}{\operatorname{D} t}.
\end{equation}
The above formula also holds for fluid $b$,
\begin{equation}\label{eq:5-eq-b}
\frac{\operatorname{D} p_a^*}{\operatorname{D} t}+\rho_a (c_a^*)^2 \operatorname{div}\bm{u}=-\frac{\rho_a (c_a^*)^2}{z_a} \frac{\operatorname{D} z_{a}}{\operatorname{D} t}.
\end{equation}
We label with superscript $\infty$ the variables under the instantaneous velocity equilibrium and pressure equilibrium.  Then $p_a^\infty =p_b^\infty=p$.
In view of stiff pressure relaxation, the following equation is obtained by subtracting  \eqref{eq:5-eq-a} from \eqref{eq:5-eq-b},
\begin{equation}
\left(\rho_a c_a^2-\rho_b c_b^2\right)^\infty \operatorname{div}\bm{u}+\sum_{k=a,b}\frac{\left(\rho_k c_k^2\right)^\infty}{z_k^\infty} \frac{\operatorname{D} z_{a}^\infty}{\operatorname{D} t}=\frac{\operatorname{D} (p_b^\infty-p_a^\infty)}{\operatorname{D} t}=0.
\end{equation}
Thus the evolutionary equation of the volume fraction becomes
\begin{equation}\label{eq:5-eq-5}
\frac{\partial z_{a}^\infty}{\partial t}+\bm{u} \cdot \nabla z_{a}^\infty= \omega^\infty \operatorname{div} \bm{u},\quad
\omega:=z_{a} z_{b} \frac{\rho_{b} c_{b}^{2}-\rho_{a} c_{a}^{2}} {\sum_{k=a,b} z_{k^{\prime}} \rho_{k} c_{k}^{2}}.
\end{equation}
Moreover, the total energy equation for the mixture is obtained by the sum of (\ref{eq:6-eq}d) and (\ref{eq:6-eq}e),
\begin{equation}\label{eq:5-eq-4}
\frac{\partial \rho E}{\partial t}+\operatorname{div}\left[(\rho E+p) \bm{u}\right]=0.
\end{equation}
Finally, (\ref{eq:6-eq}a)-(\ref{eq:6-eq}c), \eqref{eq:5-eq-5} and \eqref{eq:5-eq-4} form the reduced five-equation model written in term of conservative variables,
\begin{subequations}\label{eq:5-eq}
\begin{align}
&\frac{\partial \rho}{\partial t}+\operatorname{div}\left(\rho \bm{u}\right)=0,\\
&\frac{\partial \rho \bm{u}}{\partial t}+\operatorname{div}(\rho \bm{u} \otimes \bm{u})+\nabla p=0,\\
&\frac{\partial \rho E}{\partial t}+\operatorname{div}\left[(\rho E+p) \bm{u}\right]=0,\\
&\frac{\partial \rho \phi_{a}}{\partial t}+\operatorname{div}\left(\rho \phi_{a} \bm{u}\right)=0,\\
&\frac{\partial z_{a}}{\partial t}+\bm{u} \cdot \nabla z_{a}=\omega\operatorname{div} \bm{u},
\end{align}
\end{subequations}
where the mixture sound speed of this model obeys the Wood \cite{wood1930textbook} formula
\begin{equation}\label{eq:Wood}
\frac{1}{\rho c^2}=\sum_{k=a,b}\frac{z_k} {\rho_k c_k^2}.
\end{equation}

In the five-equation model, this non-conservative contribution $\omega\operatorname{div} \bm{u}$ makes it difficult to maintain a positive volume fraction, especially in the case of shocks and strong rarefaction waves.
As emphasized in \cite{richard_saurel_simple_2009}, numerically it is more advantageous to solve the six-equation model with stiff mechanical relaxation rather than the five-equation reduced model.
Therefore, we improve the six-equation model by involving the kinetic energy exchange and imitate the computational method of the volume fraction in the five-equation model. A new reduced model is proposed in the next section.


\subruninhead{Five-equation transport model.}

In early studies such as \cite{allaire_five-equation_2002}, the evolutionary equation for $z_a$ is a simple transport equation,
\begin{equation}\label{eq:adven-z}
\frac{\partial z_{a}}{\partial t}+\bm{u} \cdot \nabla z_{a}=0.
\end{equation}
As one can see from the above proof, this method violates the fact that the material derivatives of the entropies of the material $a$ are zero in \eqref{eq:D-Dt3}.
In fact, the additional term $\omega \operatorname{div} \bm{u}$ in the five-equation reduced model plays an important role in this aspect. 

\subruninhead{The $\gamma$-law.} 

For the polytropic gases, the  combination of \eqref{ideal_EOS} and \eqref{eq:cons-prop2} becomes
\begin{equation}
\frac{p}{\gamma-1} = \rho e = z_a \rho_a e_a+ z_b \rho_b e_b = \sum_{k=a,b}\frac{z_k p_k}{\gamma_k-1}.
\end{equation}
Under the pressure equilibration $p_a^\infty =p_b^\infty=p$, the ratio of specific heats for mixture satisfies
\begin{equation}\label{eqq}
\frac{1}{\gamma-1} = \sum_{k=a,b} \frac{z_k^\infty}{\gamma_k-1}.
\end{equation}
Then, (\ref{eq:5-eq}e) can be written in the following form
\begin{equation}\label{eq:5-eq-5-2}
\frac{\partial z_{a}^\infty}{\partial t}+\bm{u} \cdot \nabla z_{a}^\infty=  \frac{\frac{1}{\gamma_b-1}-\frac{1}{\gamma_a-1}}{\sum_{k=a,b} \frac{1}{z_k^\infty(\gamma_k-1)}}
\operatorname{div} \bm{u}.
\end{equation}
It is easy to see that, for the polytropic gases, (\ref{eq:5-eq}d) is redundant in the five-equation model \eqref{eq:5-eq}. We believe that adding the effect of the mass fraction will increase the accuracy of the model. This is achieved in the following reduced model.



\subsection{A novel reduced model}\label{sec:new-reduced}


We present a method to estimate the amount of kinetic energy exchange caused by velocity relaxation.
Under instantaneous pressure equilibrium $p_a^\infty=p$ and velocity equilibrium $\bm{u}_a^\infty=\bm{u}$, the velocity $\bm{u}_a^0$ is defined as the solution of (\ref{cons-eqns}b) in the seven-equation model without the velocity relaxation term,
\begin{equation}\label{momentum-a-eqn}
\frac{\partial \rho \phi_{a} \bm{u}_{a}^0}{\partial t}+\operatorname{div}\left(\rho \phi_{a} \bm{u} \otimes \bm{u}\right)+z_{a}^\infty \nabla p=0.
\end{equation}
Then the subtraction of  the above equation from  (\ref{cons-eqns}b), we can estimate the velocity relaxation term by this way,
\begin{equation}\label{v-relax}
\lambda \left(\bm{u}_{b}^\infty-\bm{u}_{a}^\infty\right) = \frac{\partial \rho \phi_{a} \left(\bm{u}_a^\infty-\bm{u}_a^0\right)}{\partial t} = \frac{\partial \rho \phi_{a} \left(\bm{u}-\bm{u}_a^0\right)}{\partial t},
\end{equation}
for $\lambda\rightarrow \infty$.
It is easy to see in this form that the velocity relaxation term reflects the magnitude of the kinetic energy exchange.
Formally, the mass fraction must affect the velocity relaxation and thus the result of the model.
Substituting  it into (\ref{cons-eqns}c) yields
\begin{equation}\label{energy-a-eqn0}
\frac{\partial \rho \phi_{a} E_{a}^\infty}{\partial t}+\operatorname{div}\left[\left(\rho \phi_{a} E_{a}^\infty+z_{a}^\infty p_{a}^\infty\right) \bm{u}\right]=-p \frac{\partial z_{a}^\infty}{\partial t}+\bm{u}\cdot \frac{\partial \rho \phi_{a} \left(\bm{u}-\bm{u}_a^0\right)}{\partial t}.
\end{equation}
To obtain a reduced model, a remaining question is how to express $p \partial z_a^\infty\big/\partial t$ in \eqref{energy-a-eqn0}.
We substitute the evolution equation of the volume fraction \eqref{eq:5-eq-5} into \eqref{energy-a-eqn0}, and then get the total energy equation for the material $a$,
\begin{equation}\label{energy-a-eqn}
\frac{\partial \rho \phi_{a} E_{a}^\infty}{\partial t}+\operatorname{div}\left(\rho \phi_{a} E_{a}^\infty \bm{u}\right)+z_{a}^\infty \operatorname{div}\left(p \bm{u}\right)=\bm{u}\cdot \frac{\partial \rho \phi_{a} \left(\bm{u}-\bm{u}_a^0\right)}{\partial t}-\omega^\infty p \operatorname{div} \bm{u}.
\end{equation}
Then a novel reduced model is formed by (\ref{eq:5-eq}a)-(\ref{eq:5-eq}d), (\ref{momentum-a-eqn}) and \eqref{energy-a-eqn}.
That is to say, over the control volume of the mixture, the governing equations for the mixture and the material $a$ take the form
\begin{equation}\label{Euler_eq}
\frac{\partial }{\partial t}\bm{U}+\operatorname{div} \bm{F}(\bm{U})+ z_a \operatorname{div} \bm{G}(\bm{U}) + \omega p \operatorname{div}\bm{R}(\bm{U}) = \bm{u}\cdot \frac{\partial }{\partial t}\bm{Q}(\bm{U}),
\end{equation}
with
\begin{equation*}
\bm{U}=
\begin{bmatrix}
\rho\\
\rho \bm{u}\\
\rho E\\
\rho \phi_a\\
\rho\phi_a \bm{u}_a^0\\
\rho\phi_a E_a
\end{bmatrix},\,
\bm{F}=
\begin{bmatrix}
\rho \bm{u}\\
\rho \bm{u}\otimes \bm{u} + p \bm{I}\\
(\rho E +p)\bm{u}\\
\rho\phi_a \bm{u}\\
\rho\phi_a \bm{u}\otimes\bm{u}\\
\rho\phi_a E_a \bm{u}
\end{bmatrix},\,
\bm{G}=
\begin{bmatrix}
0\\
0\\
0\\
0\\
p\bm{I}\\
p\bm{u}
\end{bmatrix},\,
\bm{R}=
\begin{bmatrix}
0\\
0\\
0\\
0\\
0\\
\bm{u}
\end{bmatrix},\,
\bm{Q}=
\begin{bmatrix}
0\\
0\\
0\\
0\\
0\\
\rho \phi_{a} \left(\bm{u}-\bm{u}_a^0\right)
\end{bmatrix}.
\end{equation*}
A system of equations symmetric with respect to \eqref{Euler_eq} can be derived for the material $b$, which is equivalent to \eqref{Euler_eq}.
For any control volume in which the flow field is continuously differentiable, the last term in the system \eqref{Euler_eq} can be written as
\begin{equation}
\frac{\partial \rho \phi_{a} \left(\bm{u}-\bm{u}_a^0\right)}{\partial t} = 
\lambda \left(\bm{u}_{b}^\infty-\bm{u}_{a}^\infty\right) = 
\Sigma^\infty - p\nabla z_a^\infty = \left(z_a^\infty-\phi_a\right) \nabla p,
\end{equation}
for $\lambda\rightarrow \infty$. However, it can be expressed by conservative variables
\begin{equation}
\rho \phi_{a} \left(\bm{u}-\bm{u}_a^0\right) = \frac{(\rho\bm{u})(\rho\phi_a)}{\rho} - \rho\phi_a \bm{u}_a^0 = \frac{\bm{U}_2\bm{U}_4}{\bm{U}_1} - \bm{U}_5,
\end{equation}
which is significant across  discontinuities (shocks).
We note that the cell average of non-conservative product $\left(z_a^\infty-\phi_a\right) \nabla p$ has no physical sense. Thus, numerical discretization of $\partial \bm{Q}/\partial t$ should be more accurate based on the conservative variables.

 In order to accurately design numerical schemes with high fidelity, we summarize that the model has the following properties:
\begin{itemize}
\item The equations for different materials are symmetric. The equations for the material $a$ in the model are formally consistent with the equations of the mixture.
\item This model reflects the exchange of kinetic energy. Hence it is favorable for the simulation of shocks near material interfaces.
\item Similar to the form of the six-equation model,  it is helpful to take account of  the energy equation rather than  the volume fraction equation to keep the positivity of the volume fraction. 
\item This model is easy to be discretized   and extended to high-order accuracy and multi-dimensional cases.
\end{itemize}

These properties inform that the corresponding numerical schemes for this model must be symmetric and the exchange of kinetic energy should be computed reasonably to ensure the positivity of the internal energy.
It is necessary to generalize the numerical schemes to higher-order  accuracy and multidimensional cases.

\section{Discretization methods}

Two-material models in the previous  are in the non-conservative form. 
The next question is how to discretize the non-conservative systems.
It is worth noting that there are many difficulties and challenges in numerical discretization for  non-conservative systems.
In the Eulerian framework, numerical solutions by non-conservative numerical schemes may tend to converge to wrong solutions \cite{hou_why_1994}, providing incorrect partition of internal energies or shock wave velocity in the shock layer.


The first four equations in the five-equation model \eqref{eq:5-eq} are conservative Euler equations, forming the four-equation model \cite{abgrall_computations_2001}.
For the four-equation model, the most common approach for the closure is the assumption of temperature equilibrium.
There are numerical difficulties due to nonphysical oscillations generated at material interfaces when conservative schemes are applied to the four-equation model.

Although it is not appropriate to use the conservative four-equation model, it is better to consider a model similar to the conservative Euler system in form.
For this purpose, we propose a numerical algorithm for the novel reduced model based on the Godunov-type method,
to correctly simulate the numerical phenomena of the compressible flow involving material interfaces, shock waves and rarefaction waves.
Before that, let us take a look at some of the existing numerical methods for the two-material models in the previous section.

\subsection{Conventional Eulerian methods for the five-equation model}

In this section, we begin by introducing a finite volume method, i.e. the conventional Godunov-type scheme, for solving the five-equation models over the Eulerian grid.

\subsubsection{The Riemann problem}

The Riemann problem is the building block of Godunov-type schemes. 
For the 1-D five-equation model \eqref{eq:5-eq}, the initial data at time $t=t_n$ is assumed to be piece-wise constant and the discontinuity is shifted to $t=0$ by the Galilean invariance,
\begin{equation*}
\bm{V}(x,t=0)=\left\{
\begin{aligned}
&\bm{V}_L, & x<x_{0},\\
&\bm{V}_R, & x>x_{0},
\end{aligned}
\right.
\end{equation*}
where $\bm{V}=[\rho, \rho \bm{u}, \rho E, \rho \phi_a, z_a]^\top$, $\bm{V}_L$ and $\bm{V}_R$ are constant states on both sides of certain position $x=x_{0}$. The solution of this Riemann problem, denoted as $\textbf{RP}(\bm{V}_L,\bm{V}_R)$, has self-similarity,
\begin{equation*}
\bm{V}(x,t)=\textbf{v}(\xi),\quad\  \xi=\frac{x-x_{0}}{t},\ t>0. 
\end{equation*}
This solution consists of a left-facing shock or rarefaction wave $\mathcal{W}_0^-$; a right-facing shock or rarefaction wave $\mathcal{W}_0^+$ and a contact discontinuity  $\mathcal{C}$ with the velocity $u_0^*$.

\subsubsection{Conventional Godunov-type schemes}

In the 1-D case, the computational domain $[0, L]$ is divided into $M$ fixed grid cells $I_i=\left[x_{i-\frac12},x_{i+\frac12}\right], i=1,2,\ldots,M$,   with the grid size $\Delta x=x_{i+\frac12}-x_{i-\frac12}=L/M$, the cell interface $x_{i+\frac12}=i\Delta x$,  and the cell center $x_i=\left(i-\frac12\right)\Delta x$.
The conservative equations of the five-equation system \eqref{eq:5-eq} is updated from $t_n$ to $t_{n+1}=t_n+\Delta t$ by the conventional Godunov scheme
\begin{equation}
\bm{W}_i^{n+1} =  \bm{W}_i^{n} - \Lambda \left[F\left(\bm{W}_{i+\frac12}^{n}\right) - F\left(\bm{W}_{i-\frac12}^{n}\right)\right],
\end{equation}
where $\Lambda=\Delta t/\Delta x$, $\bm{W}=\bm{U}^{(4)}=[\rho, \rho \bm{u}, \rho E, \rho \phi_a]$ is the first four variables of $\bm{U}$ and
\begin{equation}
\bm{W}_{i+\frac12}^{n} = \textbf{v}^{(4)}(\xi=0),\quad\  \xi=\frac{x-x_{i+\frac12}}{t-t_n}
\end{equation}
in the numerical flux, where the superscript $(4)$ refers to the four components. The numerical fluxes can be evaluated by the exact Riemann solver or approximate Riemann solvers, such as an acoustic Riemann solver \cite{murrone_five_2005}.
In the five-equation transport model, the advenction equation \eqref{eq:adven-z} of the volume fraction 
\begin{equation}
\frac{\partial z_{a}}{\partial t}+\operatorname{div} (z_{a}\bm{u}) = z_a \operatorname{div} \bm{u}
\end{equation}
is update by a Godunov-type scheme
\begin{equation}
(z_a)_i^{n+1} =  (z_a)_i^{n} - \Lambda \left[\left(z_a u\right)_{i+\frac12}^{n}-\left(z_a u\right)_{i-\frac12}^{n}-(z_a)_i^{n}\left(u_{i+\frac12}^{n}-u_{i-\frac12}^{n}\right)\right],
\end{equation}
which easily preserves positivity of the volume fraction.
In the five-equation reduced model, the evolution equation \eqref{eq:5-eq-5} of the volume fraction 
\begin{equation}
\frac{\partial z_{a}}{\partial t}+\operatorname{div} (z_{a}\bm{u})=(\omega + z_a)\operatorname{div} \bm{u}
\end{equation}
is updated by
\begin{equation}
(z_a)_i^{n+1} =  (z_a)_i^{n} - \Lambda \left[\left(z_a u\right)_{i+\frac12}^{n}-\left(z_a u\right)_{i-\frac12}^{n}-(\omega + z_a)_{i}^{n}\left(u_{i+\frac12}^{n}-u_{i-\frac12}^{n}\right)\right].
\end{equation}
The additional term $\omega_{i}^{n}\left(u_{i+\frac12}^{n}-u_{i-\frac12}^{n}\right)$ makes it difficult to maintain positivity of the volume fraction.
Let us refer a numerical example to illustrate troubles of these Godunov-type schemes.

\begin{example}[Epoxy-spinel mixture shock tube problem]\label{ex-1}

A shock tube in $[0,1]$ consists of two segments filled with the same mixture of epoxy and spinel, and separated by an interface at $x=0.6$.
The specific initial data for this example can be found in \cite{saurel_relaxation-projection_2007}, including two cases of moderate pressure ratio and extreme pressure ratio.
\end{example}

\begin{figure}[htb]
\centering
\includegraphics[width=9cm]{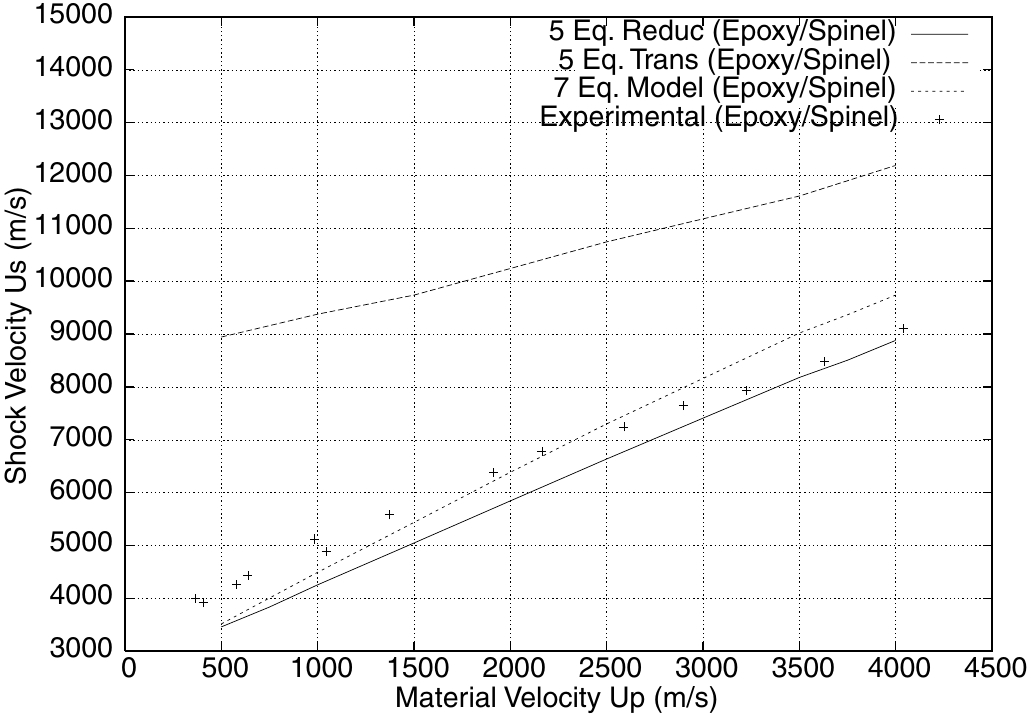}
\caption{Comparison of numerical results of variable impact velocity $U_\mathrm{p}$ for the five-equation transport model, the five-equation reduced model and the seven-equation model, with experimental data.}\label{fig:murrone1}
\end{figure}

For the epoxy-spinel mixture shock tube problem with moderate pressure ratio, 
by computing several cases with variable impact velocities, we compare 
numerical results of the five-equation transport model, five-equation reduced model and seven-equation model in Fig.~\ref{fig:murrone1}.
We find that the result of the five-equation reduced model is close to that of the seven-equation model, but far from that of the five-equation transport model.
It shows that the additional term in the five-equation reduced model has an obvious effect, and the five-equation transport model is not suitable for compressible multi-material flows.

For the five-equation model,  the Wood sound speed of the mixture \eqref{eq:Wood}  has a non-monotonic variation with volume fraction in the numerical diffusion zone of material interfaces.
As explained in \cite{saurel_relaxation-projection_2007}, it may result in the presence of two sonic points in this region and difficulties for the Riemann problem resolution.
The non-monotonic behavior of the sound speed comes from the equilibrium condition $p_a^\infty = p_b^\infty$, we can avoid it by solving a non-equilibrium model with relaxation.
Another way to circumvent this difficulty is to use Lagrange-projection method for the five-equation model since the Lagrange step does not need the complete Riemann problem resolution.

\subsection{Lagrange-projection method of the five-equation model}

\begin{figure}[htb]
\centering
\includegraphics[width=8cm]{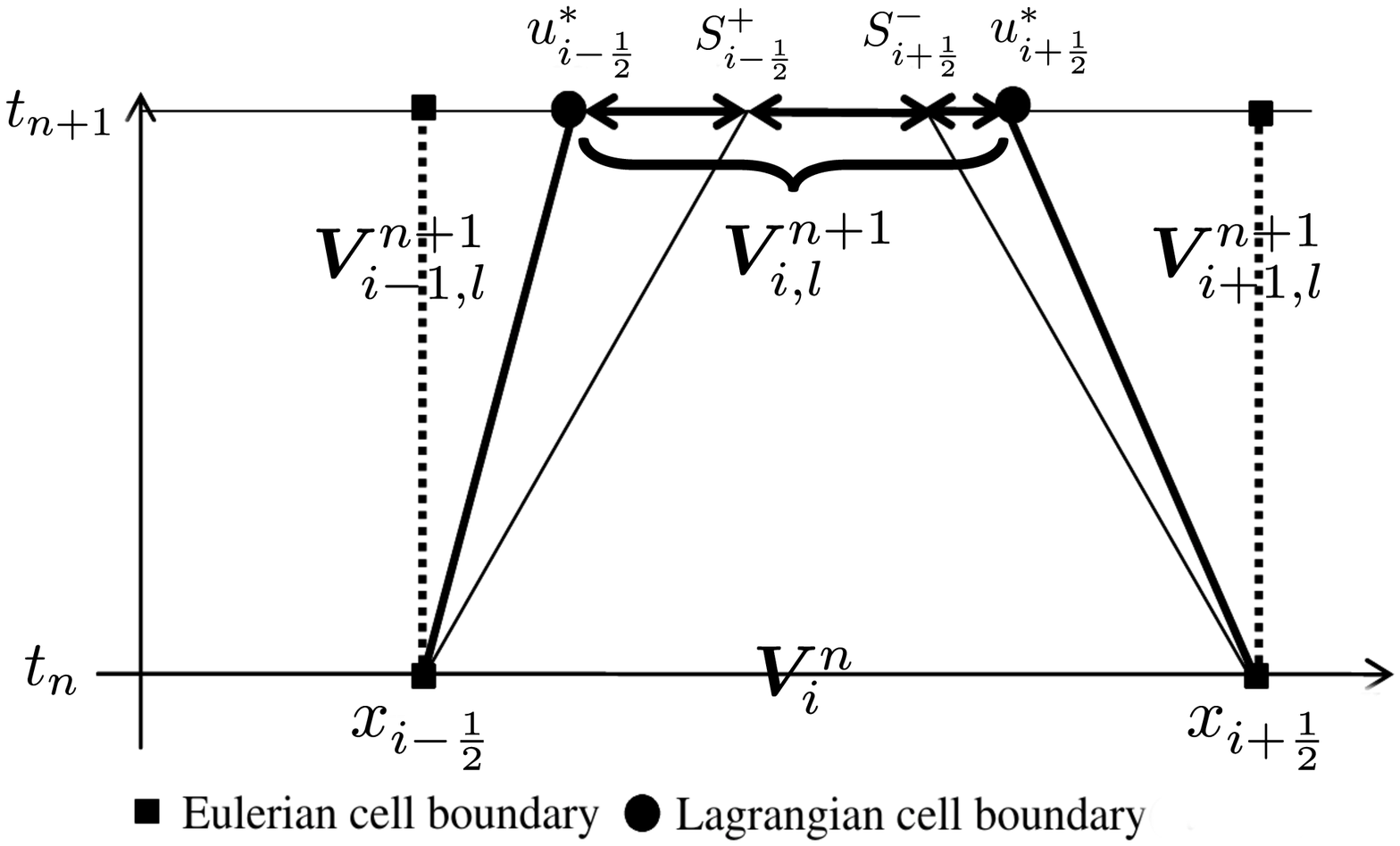}
\caption{Schematic diagram of the Lagrange-projection method.}\label{fig:L-p-schematic}
\end{figure}

As shown in the procedure diagram of Fig.~\ref{fig:L-p-schematic}, the Lagrange-projection method consists of two steps.
The first step rests in the solution of the governing equations \eqref{eq:5-eq} in terms of  Lagrange coordinates. The second step corresponds to the projection of the solution onto the Eulerian grid.

\subsubsection{Lagrange step}
	
For the Lagrangian computational cell $J_i(t) = \left[x_{i-\frac12}(t), x_{i+\frac12}(t)\right]$, the cell boundary $x_{i+\frac12}$ moves with the velocity $u_{i+\frac12}^*$. Then 
the grid size of $J_i(t_{n+1})$ is $\Delta x_i^{n+1} = \Delta x_i^n + \Delta t \left(u_{i+\frac12}^*-u_{i-\frac12}^*\right)$, where $\Delta x_i^n=x_{i+\frac12}(t_n)-x_{i-\frac12}(t_n)$.
Then, the cell average state on $J_i(t_{n+1})$ is obtained by
\begin{equation}
\bm{V}_{i,l}^{n+1}= \sum_{j=1}^3\beta_j \bm{V}_{j}^{*}
\end{equation}
with three piece-wise constant states separated by left-facing and right-facing waves $S^\mp_{i\pm \frac12}$
\begin{equation}
\bm{V}_{1}^{*}=\bm{V}_{R,i-\frac12}^{*},\quad
\bm{V}_{3}^{*}=\bm{V}_{L,i+\frac12}^{*},\quad
\bm{V}_{2}^{*}=\bm{V}_i^{n},
\end{equation}
where the subscripts $L$ and $R$ refer to the left and right states in the Riemann problem, respectively, and the corresponding volume fractions in these Lagrangian subcells are
\begin{equation}
\beta_1 = \frac{\Delta t}{\Delta x_i^{n+1}}\left(S_{i-\frac12}^+ - u_{i-\frac12}^*\right),\quad
\beta_3 = \frac{\Delta t}{\Delta x_i^{n+1}}\left(u_{i+\frac12}^* - S_{i+\frac12}^-\right),\quad
\beta_2 = 1-\beta_1-\beta_3.
\end{equation}

\subsubsection{Projection on the Eulerian grid}

\begin{figure}[htb]
\centering
\includegraphics[width=11cm]{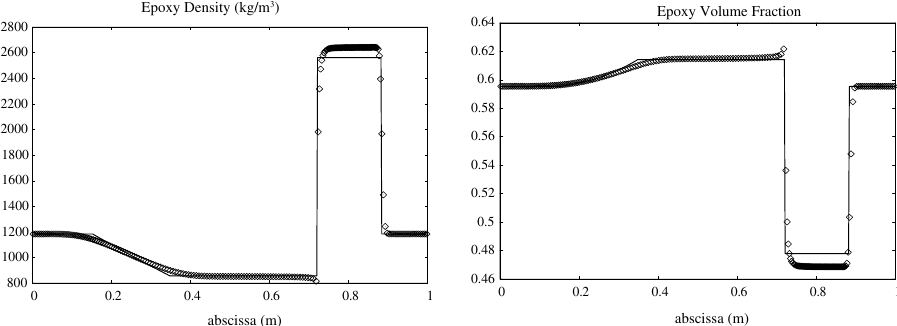}
\caption{Comparison of numerical results by the Lagrange-projection (conventional Godunov-type) method ($\lozenge$) and the exact solutions (--) for the five-equation model.}\label{fig:L-p-compare}
\end{figure}

The numerical solution in the Eulerian cell $I_i$ is obtained by the projection of the Lagrangian cell average,
\begin{equation}\label{eq:project-Lag}
\bm{V}_{i}^{n+1}=\frac{1}{\Delta x} \sum_{j=i-1}^{i+1}L_j \bm{V}_{j,l}^{n+1},
\end{equation}
where the sizes of three sub-volumes, separated by Lagrangian boundaries, are
\begin{equation}
L_1=\max\left(0,u_{i-\frac12}^*\right)\Delta t,\quad
L_3=-\min\left(0,u_{i+\frac12}^*\right)\Delta t,\quad
L_2=\Delta x-L_1-L_3.
\end{equation}
For the polytropic gases, the Lagrange-projection method is equivalent to the conventional Godunov-type scheme.
Again, we utilize Example \ref{ex-1} to illustrate the difficulty of the Lagrange-projection method.
For the five-equation model, the numerical results of the Lagrange-projection method are compared with the exact solutions in Fig.~\ref{fig:L-p-compare}.
The numerical solutions of the density and volume fraction for epoxy are not in conformity with the exact solutions in vicinity of the shock front.
The errors between the two solutions remain the same after the mesh is refined.
This example illustrates that numerical solutions by  the conventional Godunov-type scheme is difficult to converge to the exact solution for strong multi-material shocks.

We note that the volume fraction is not a conservative variable.
Hence, the Eulerian cell average of the volume fraction in \eqref{eq:project-Lag} is not physically correct.
It results in an incorrect partition of the internal energies (or entropies) between two materials \cite{saurel_relaxation-projection_2007} for strong multi-material shocks.
Hence, in order to achieve more correct partition of the shock energy, we move on to  another method, i.e. solving a non-equilibrium model with relaxation.

\subsection{Fractional step method for the non-equilibrium model}

As summarized in the above discussion on discretizing the five-equation reduced model for  the presence of multi-material shocks or strong expansion waves, it is a difficult issue about the   convergence of the numerical solutions and the positivity preserving of the volume fraction.
To overcome these difficulties, the mechanical equilibrium assumption is relaxed and non-equilibrium models are utilized.
The non-equilibrium models, composed of the seven-equation model and the six-equation model, are non-conservative hyperbolic models involving relaxation terms.
The numerical approximation of the non-equilibrium models is implemented by a fractional step technique with two steps: hyperbolic evolution and mechanical relaxation.
Only considering the pressure relaxation in the mechanical relaxation step, we introduce the fractional step method based on the six-equation model \eqref{eq:6-eq-compact}.


As shown concretely in \cite{pelanti_mixture-energy-consistent_2014}, the initial value (IV) problem of the six-equation system \eqref{eq:6-eq-compact} is solve by the hyperbolic evolution step
\begin{equation}\label{eq:hyperbolic}
\left.\begin{aligned}
&\frac{\partial \mathcal{U}}{\partial t}+\operatorname{div}\mathcal{F}(\mathcal{U})+\bm{u}\cdot \mathcal{S}(\mathcal{U}) = 0\\
\mbox{IV: }&\mathcal{U}(x,t_n)=\mathcal{U}^n
\end{aligned}\right\}\Longrightarrow
\mathcal{U}^*,
\end{equation}
and the pressure relaxation step
\begin{equation}\label{eq:ODE}
\left.\begin{aligned}
&\frac{\partial \mathcal{U}}{\partial t} = \mu \mathcal{R}(\mathcal{U})\\
\mbox{IV: }&\mathcal{U}(x,t_n)=\mathcal{U}^*
\end{aligned}\right\}\Longrightarrow
\mathcal{U}^{n+1}.
\end{equation}
Here the solution for the first step is $\mathcal{U}^*$ and the solution for the second step is $\mathcal{U}^{n+1}$.

\subsubsection{Hyperbolic evolution step}

We begin by numerically solving the homogeneous hyperbolic part of the six-equation system in \eqref{eq:hyperbolic}.
For 1-D cases, this hyperbolic system can be written in a quasilinear form,
\begin{equation}\label{eq:quasi-linear}
\frac{\partial \mathcal{U}}{\partial t}+ \mathcal{A}(\mathcal{U})\frac{\partial \mathcal{U}}{\partial x} = 0.
\end{equation}
For example, the wave-propagation scheme in \cite{pelanti_mixture-energy-consistent_2014} is employed by integrating \eqref{eq:quasi-linear} from $t_n$ to $t_{n+1}$,
\begin{equation}
\mathcal{U}_i^* =  \mathcal{U}_i^{n} - 
\Lambda \left[\mathcal{A}^+\Delta \mathcal{U}_{i-\frac12}^{n} + \mathcal{A}^-\Delta \mathcal{U}_{i+\frac12}^{n}\right],
\end{equation}
with the fluctuations at the cell interface $x_{i-\frac12}$ 
\begin{equation}
\mathcal{A}^{-} \Delta \mathcal{U}_{i-\frac12}^{n} = \sum_{p=1}^{6}\left(\lambda_{i-\frac12}^{p}\right)^{-} \mathcal{W}_{i-\frac12}^{p},\quad
\mathcal{A}^{+} \Delta \mathcal{U}_{i-\frac12}^{n} = \sum_{n=1}^{6}\left(\lambda_{i-\frac12}^{p}\right)^{+} \mathcal{W}_{i-\frac12}^{p}
\end{equation}
determined by solving the Riemann problems, where the jump $\Delta \mathcal{U}_{i-\frac12}^{n} = \mathcal{U}_{i}^n-\mathcal{U}_{i-1}^n$ decomposed into a set of waves $\mathcal{W}_{i-\frac12}^{p}$ propagating with speeds $\lambda_{i-\frac12}^{p}$ for $p = 1, 2, \ldots, 6$ so that
\begin{equation}
\Delta \mathcal{U}_{i-\frac12}^{n} =\sum_{p=1}^{6} \mathcal{W}_{i-\frac12}^{p},\quad
\mathcal{F}\left(\mathcal{U}_{i}^n\right)-\mathcal{F}\left(\mathcal{U}_{i-1}^n\right)+\Delta x\left(\bm{u} \cdot \mathcal{S}\right)_{i-\frac12}^{n} =\sum_{p=1}^{6} \lambda_{i-\frac12}^{p} \mathcal{W}_{i-\frac12}^{p},
\end{equation}
and $\lambda^{+}=\max (\lambda, 0), \lambda^{-}=\min (\lambda, 0)$.
This wave-propagation scheme is a kind of Godunov-type scheme.
As already mentioned above, the Godunov-type scheme guarantees the positivity of the volume fraction during the hyperbolic evolution step.

\subsubsection{Stiff pressure relaxation step}


In the pressure relaxation step, we solve the system of ordinary differential equations \eqref{eq:ODE} in the limit $\mu\rightarrow \infty$ for relaxing the pressures of two materials to an equilibrium value.
In this system, the energy of the material $k$ satisfies the equation
\begin{equation}\label{eq:relax-stiff-E}
\frac{\partial \rho \phi_{k} E_{k}}{\partial t}= - p_\mathrm{I} \frac{\partial z_{k}}{\partial t},
\end{equation}
If we integrate \eqref{eq:relax-stiff-E} from $t_n$ to $t_{n+1}$ to get
\begin{equation}
\rho\phi_{k}\left(E_{k}^{n+1}-E_{k}^{*}\right)+\int_{t_{n}}^{t_{n+1}} p_\mathrm{I} \frac{\partial z_k}{\partial t} \mathrm{d} t=0
\end{equation}
and approximate $p_\mathrm{I}$ as a constant $p_\mathrm{I} \doteq (p^{n+1}+p^*)/2$ (cf. \cite{lallemand_pressure_2005,pelanti_mixture-energy-consistent_2014}),  
the difference in internal energy after and before pressure relaxation is
\begin{equation}\label{eq:p_I_approx}
\begin{aligned}
& \left(\rho \phi_{k} e_{k}\right)^{n+1}-\left(\rho \phi_{k} e_{k}\right)^{*} = \left(\rho \phi_{k} E_{k}\right)^{n+1}-\left(\rho \phi_{k} E_{k}\right)^{*}\\
=& -\int_{t_{n}}^{t_{n+1}} p_\mathrm{I} \frac{\partial z_{k}}{\partial t}dt = -\int_{z_k^*}^{z_k^{n+1}} p_\mathrm{I}  dz_k
\doteq -\frac{p^{n+1}+p^{*}}{2}\left(z_{k}^{n+1}-z_{k}^{*}\right),
\end{aligned}
\end{equation}

\begin{figure}[htb]
\centering
\includegraphics[width=11cm]{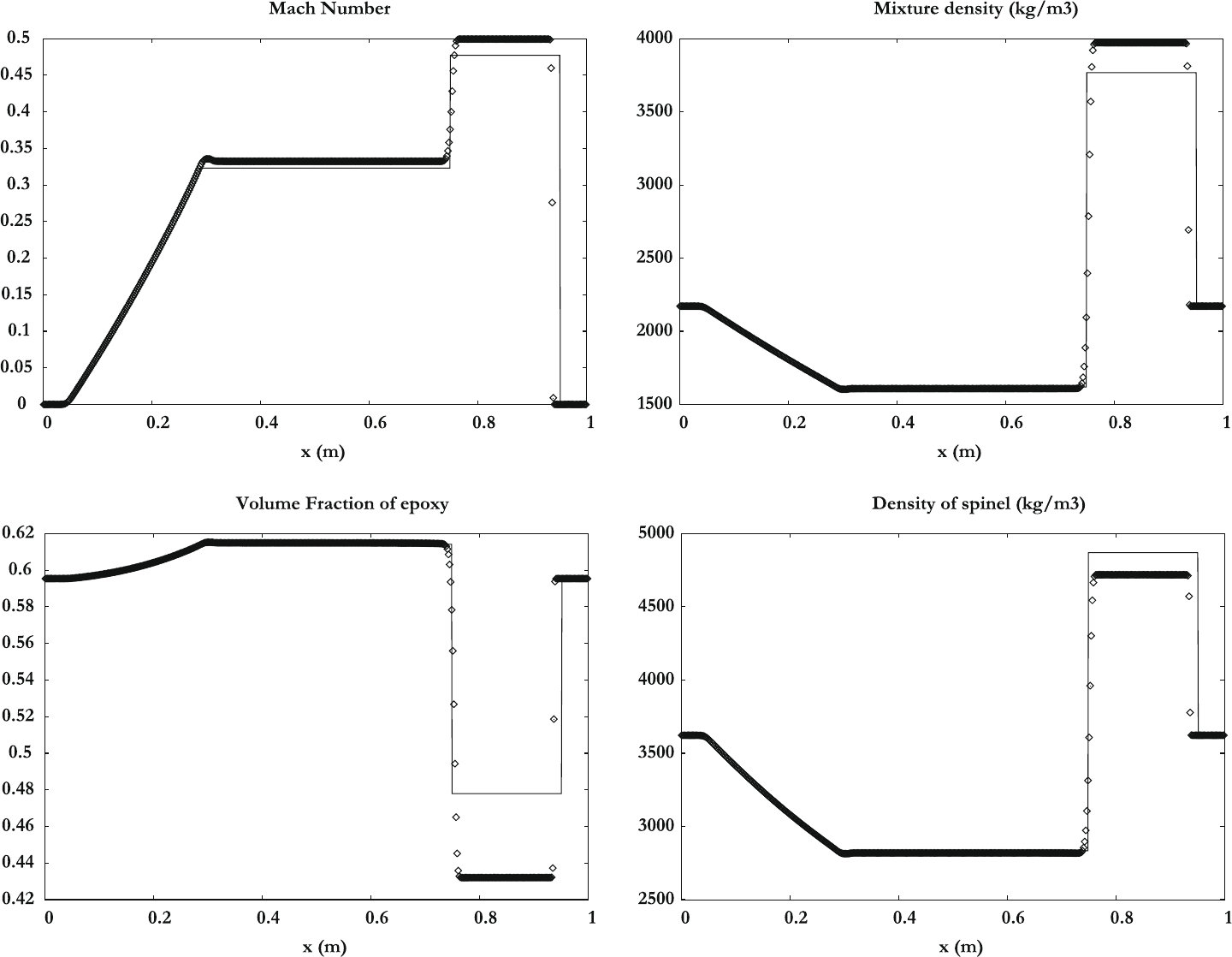}
\caption{Comparison of numerical results of the six-equation model with the present relaxation method ($\lozenge$) and the exact solution of the five-equation model (--).}\label{fig:extreme}
\end{figure}

We first test the Example \ref{ex-1} with extreme pressure ratio, and compare the numerical results of the six-equation model with the present relaxation method and the exact solution of the five-equation model.
In fact, numerical results of the six-equation model are very sensitive since the model contains a large number of waves.
As shown in Fig.~\ref{fig:extreme}, since the multi-material shock is very strong, there is a big difference between the numerical solution and the exact solution.
It reflects that the incorrect partition of the internal energies between two materials in the shock layer also appears in the six-equation model.
For the further comparison of  the effects of the relaxation term on the seven-equation model, we pay attention to another example in the following.

\begin{figure}[htb]
\centering
\includegraphics[width=11cm]{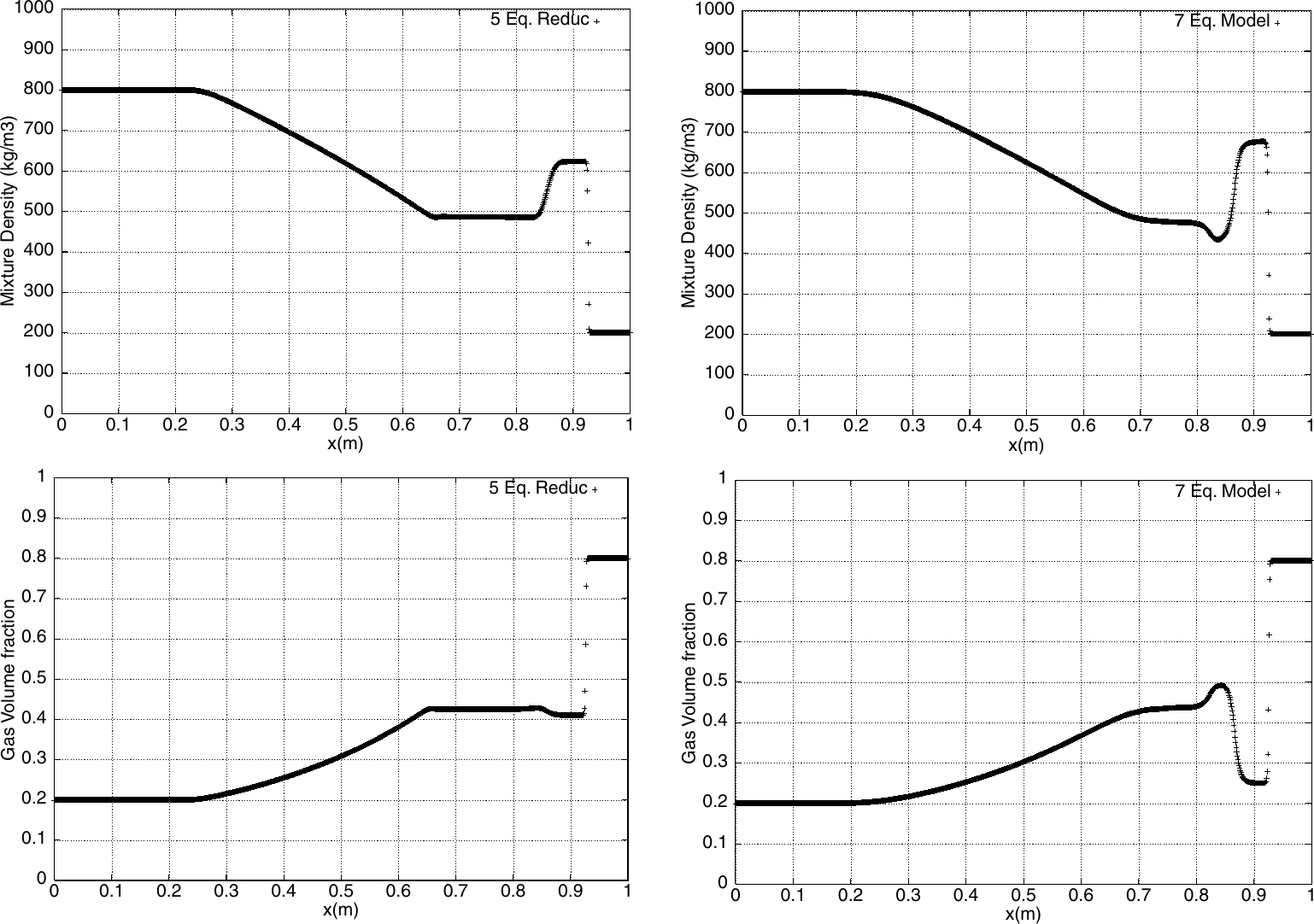}
\caption{Comparison of numerical results of the five-equation reduced model (left) and the seven-equation model (right) in a two-phase flow problem.}\label{fig:m2}
\end{figure}

\begin{example}[A two-phase flow problem]\label{ex-2}

This is the second two-phase flow problem in Section 5.2.2 of \cite{murrone_five_2005}. 
For this problem, Fig.~\ref{fig:m2} compares the numerical results of the five-equation reduced model  by the conventional Godunov-type schemes with those for the seven-equation model by the fractional step method.
There are significant differences between the numerical results of these two models, showing oscillations near the material interfaces. However, it is hard to say which is more accurate.
This suggests that it is not easy to simulate an non-equilibrium model correctly by the fractional step method.
\end{example}

\subsection{Finite volume methods for the novel reduced model}


Since there are many difficulties in the numerical discretization of the five-equation model or non-equilibrium models, we try to design numerical methods based on our novel reduced model.
This model includes an equation of mass fraction and a one-phase energy equation coupling with the Euler equations.
In order to simulate kinetic energy more accurately, we add a one-phase momentum equation for correction.
As far as the strong multi-material shocks is concerned, the exchange of kinetic energy is well worth investigaing. 

A benefit of this model is that its form is similar to that of the conservative Euler system, which allows to suit for the conventional Godunov-type scheme.
A  second order accurate extension of the conventional Godunov-type is made  using the space-time coupled generalized Riemann problem (GRP) solver \cite{ben-artzi_direct_2006,ben-artzi_generalized_2003}, which was originally proposed in \cite{BF_1984} with extension to combustion models, e.g. \cite{ben-artzi_application_1986,ben-artzi_computation_1990} and general hyperbolic balance laws \cite{Li-2}.  The reason of making this choice is the inclusion of thermodynamics into the scheme \cite{li_thermodynamical_2017} that is important in the simulation of compressible multi-material problems.  


An important issue in this type of numerical methods is related to the volume fraction positivity in the presence of shocks and even in the presence of strong rarefaction waves. Indeed, when dealing with liquid-gas mixtures, for example, the liquid compressibility is so weak that the pressure tends to be negative, resulting in computational failure in the gas sound speed computation. Such situation occurs frequently in cavitation test problems.
For this numerical method of the novel reduced model, we ameliorate this issues by simulating the exchange of kinetic energy more accurately.
Next, we introduce the concrete implementation process.

\subsubsection{2-D conventional Godunov-type scheme}

\begin{figure}[ht]
\centering
\includegraphics[width=0.35 \linewidth]{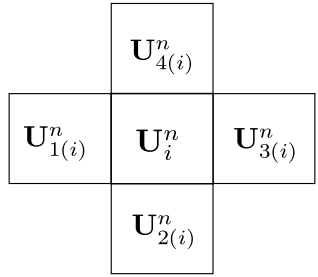}
\caption{\label{Rec}Rectangular cells and the distribution of the solution}
\end{figure}

We discretize the governing equations \eqref{Euler_eq} with a cell-centered finite-volume scheme over a two-dimensional (2-D) computational domain  divided into a set of polygonal cells $\{\Omega_i\}$. The integral average of the solution vector $\bm{U}(\bm{x},t_n)$, $\bm{x} =(x,y)$ over the cell $\Omega_i$ at time $t_n$ is given by $\bm{U}_i^n$. Taking rectangular cells as an example, we denote $\bm{U}_{j(i)}^n$ as the integral average over the $j$-th adjacent cell $\Omega_{j(i)}$ of $\Omega_i$, as shown in Fig.~\ref{Rec}.  
Since the volume fraction $z_a$ in $\Omega_i$ is expressed by solving the pressure equilibrium condition
\begin{equation}\label{Z_a_update}
p(e_a,\rho_a)=p(e_b,\rho_b)
\end{equation}
with $\rho_k=\rho\phi_k/z_k$, the vector $\bm{U}_i^n$ can be converted to $\bm{V}_i^n=\bm{V}(\bm{U}_i^n)$.
The Riemann problem $\textbf{RP}\left(\bm{V}_i^n,\bm{V}_{j(i)}^n\right)$ is solved for the five-equation system \eqref{eq:5-eq} along the unit outward normal vector $\bm n_j$  of the $j$-th boundary, and the corresponding Riemann solution is denoted by $\bm{V}_{i,j}^n$.
In the Riemann solution $\bm{V}_{i,j}^n$, the mass fraction and volume fraction are determined as
\begin{equation*}
\zeta_{i,j}^n=\left\{
\begin{aligned}
&\zeta_{i}^n, & \mbox{if }\bm{u}_{i,j}^n\cdot \bm{n}_j>0,\\
&\zeta_{j(i)}^n, & \mbox{otherwise,}
\end{aligned}
\right.
\end{equation*}
for $\zeta=\phi_a, z_a$.
Detailed process of the computation for the Riemann solution can be implemented by an acoustic approximate solver \cite{murrone_five_2005} or the exact Riemann solver \cite{ben-artzi_generalized_2003}. 

Approximating $\bm{u}$ in the last term of the five-equation system \eqref{eq:5-eq} as a constant $\bm{u}=(\bm{u}_i^n+\bm{u}_i^{n+1})/2$ (similar to the approximation for $p_\mathrm{I}$ in \eqref{eq:p_I_approx}), the finite-volume scheme with the Godunov fluxes is given by
\begin{equation}\label{FV}
\begin{aligned}
\bm{U}_i^{n+1} =\, & \bm{U}_i^{n} - \sum_{j=1}^4 \Lambda_j^i\left[\widetilde{\bm{F}}_j\left(\bm{U}_{i,j}^n\right)+(z_a)_i^n\widetilde{\bm{G}}_j\left(\bm{U}_{i,j}^n\right)+\omega_i^n p_i^n\widetilde{\bm{R}}_j\left(\bm{U}_{i,j}^n\right)\right]\\
& + \frac{\bm{u}_i^n+\bm{u}_i^{n+1}}{2}\left(\bm{Q}_i^{n+1}-\bm{Q}_i^{n}\right),
\end{aligned}
\end{equation}
where $\Lambda_j^i=\Delta t\, L_j/|\Omega_i|$, $L_j$ is the length of the $j$-th boundary of the cell $\Omega_i$, $|\Omega_i|$ is the volume of $\Omega_i$, $\bm{U}_{i,j}^n=\bm{U}(\bm{V}_{i,j}^n)$ and $\widetilde{(\bullet)}_j=(\bullet)\cdot \bm{n}_j$. 


We use the last two equations in \eqref{FV} to update the momentum and energy of material $a$.
Since the instantaneous pressure equilibrium hypothesis implies $\bm{Q}_i^{n}=\bm{0}$, the last term of $(\bm{u}_i^n+\bm{u}_i^{n+1})/2\cdot\left(\bm{Q}_i^{n+1}-\bm{Q}_i^{n}\right)$ in \eqref{FV} is
\begin{equation*}
(\Delta E_K)_i^{n+1} := \frac{\bm{u}_i^n+\bm{u}_i^{n+1}}{2}\cdot\left((\rho\phi_a\bm{u})_i^{n+1}-\left(\rho\phi_a\bm{u}_a^0\right)_i^{n+1}\right),
\end{equation*}
which is used to approximate the exchange of kinetic energy from material $b$ to material $a$ in the process of instantaneous velocity equilibrium.
After the instantaneous velocity equilibrium, the velocity of material $a$ goes from $(\bm{u}_a^0)_i^{n+1}$ to $\bm{u}_i^{n+1}$.
In addition, according to \eqref{Z_a_update}, the positivity of volume fractions is guaranteed as long as the internal energy of each material is positive.


Overall, we obtain a conventional Godunov-type scheme for the novel reduced model in two dimensions.
Such a scheme is termed as the energy-splitting  Godunov scheme ({\em ES-Godunov} for short) in \cite{lei_non-oscillatory_2018}.

\subsubsection{Second-order accurate GRP extension}

We make a second-order accurate extension of ES-Goduov  by using a 2-D generalized Riemann problem (GRP) solver  \cite{BF_1984, ben-artzi_generalized_2003,ben-artzi_direct_2006,Li-2} ({\em ES-GRP} for short).
In each cell, we project the solution vector $\bm{V}$ into the space of piecewise linear functions 
\begin{equation}
\bm{V}_i^n(\bm{x})=\bm{V}_i^n+\bm{\sigma}_i^n\cdot (\bm{x}-\bm{x}_i), 
\label{linear-data}
\end{equation}
where $\bm{\sigma}_i^n$ is the gradient of $\bm{V}$ inside the cell $\Omega_i$ at time $t=t_n$, and $\bm{x}_i$ is the centroid of $\Omega_i$.
The quasi 1-D generalized Riemann problem $\textbf{GRP}\left(\bm{V}_i^n(\bm{x}),\bm{V}_{j(i)}^n(\bm{x})\right)$ 
is solved for the five-equation system \eqref{eq:5-eq} at the center $\bm{x}_{i,j}$ of the $j$-th boundary with accuracy of second order, and then the associated Riemann solution $\bm{V}_{i,j}^n$ and the corresponding temporal derivative $\left(\partial \bm{V}/\partial t\right)_{i,j}^n$ are determined.

Referring to the conventional Godunov-type scheme \eqref{FV}, the two-dimensional finite-volume GRP scheme for \eqref{Euler_eq} is written as
\begin{equation}\label{GRP_scheme}
\begin{aligned}
\bm{U}_i^{n+1} =\, & \bm{U}_i^{n} - \sum_{j=1}^4 \Lambda_j^i \left[\widetilde{\bm{F}}_j\left(\bm{U}_{i,j}^{n+\frac{1}{2}}\right)+(z_a)_i^{n+\frac{1}{2}}\widetilde{\bm{G}}_j\left(\bm{U}_{i,j}^{n+\frac{1}{2}}\right)+\omega_i^{n+\frac12} p_i^{n+\frac12}\widetilde{\bm{R}}_j\left(\bm{U}_{i,j}^{n+\frac12}\right)\right]\\
& + \frac{u_i^n+u_i^{n+1}}{2}\left(\bm{Q}_i^{n+1}-\bm{Q}_i^{n}\right),
\end{aligned}
\end{equation}
where the mid-point value $\bm{U}_{i,j}^{n+\frac12}=\bm{U}\left(\bm{V}_{i,j}^{n+\frac12}\right)$ is determined by
\begin{equation}
\bm{V}_{i,j}^{n+\frac{1}{2}}=\bm{V}_{i,j}^n+\frac{\Delta t}{2} \left(\frac{\partial \bm{V}}{\partial t}\right)_{i,j}^n.
\end{equation}
Consistent with the whole computational procedure of the conventional Godunov-type scheme, a second-order GRP scheme for the novel reduced model in two dimensions is obtained.
The Abgrall's criterion of this second-order scheme was proof in \cite{lei_non-oscillatory_2018}, which requires uniform velocity and pressure be preserved for multi-material flows.
Therefore, this type of scheme has the non-oscillatory property.

\subruninhead{The $\gamma$-law.}

For the polytropic gases, the volume fraction of material $a$ in $\Omega_i$ is expressed as
\begin{equation*}
(z_a)_i^{n}=\frac{(\rho \phi_a e_a)_i^{n}(\gamma_a-1)}{\sum_{k=a,b}(\rho \phi_k e_k)_i^{n}(\gamma_k-1)}
\end{equation*}
through \eqref{Z_a_update}.  The effective ratio of specific heats in $\Omega_i$ is calculated as 
\begin{equation*}
\gamma_i^n=\frac{\sum_{k=a,b}(\rho \phi_k e_k)_i^{n} \gamma_k}{(\rho e)_i^n}
\end{equation*}
by using \eqref{gamma}, the mid-point value of the ratio of specific heats $\gamma_{i,j}^{n+\frac 12}$ on cell interfaces is given by 
\begin{equation}
\frac{1}{\gamma_{i,j}^{n+\frac{1}{2}}-1}=\sum_{k=a,b} \frac{z_{k,i,j}^{n+\frac{1}{2}}}{\gamma_k-1},
\end{equation}
and the last component of $\widetilde{\bm{F}}_j\left(\bm{U}_{i,j}^{n+\frac{1}{2}}\right)$ in \eqref{GRP_scheme} is evaluated by
\begin{equation*}
\left(\rho\phi_a E_a\right)_{i,j}^{n+\frac{1}{2}} \bm{u}_{i,j}^{n+\frac{1}{2}}\cdot \bm{n}_j = \left(\frac{z_a p}{\gamma_a-1}+\frac{1}{2}\rho\phi_a \bm{u}^2\right)_{i,j}^{n+\frac{1}{2}} \bm{u}_{i,j}^{n+\frac{1}{2}}\cdot \bm{n}_j.
\end{equation*} 

For readers' convenience and completeness of presentation, we put the GRP solver in Appendix. 

\section{Numerical Results}\label{Sec:num_res}

Some numerical results of the novel reduced model by using the current Godunov-type schemes are presented in this section.
We hope that these numerical results can reflect the effectiveness of the schemes to guarantee the positivity of volume fractions and correct simulation of multi-material shocks.
The kinetic energy exchange  are considered to guarantee the positive internal energy of each material numerically and thus the positive volume fraction.

We abbreviate 
{\em Is-Godunov} for the Godunov scheme of a 4-equation model with the isothermal hypothesis \cite{larrouturou_how_1991,quirk_dynamics_1996}, 
{\em UPV-Godunov} for the Is-Godunov with the energy correction based on a UPV flow \cite{banks_high-resolution_2007}, in addition to  the abbreviations:  {\em ES-Godunov} and {\em ES-GRP}.

The first numerical example is proposed to show the influence of the kinetic energy exchange terms on the positivity of volume fractions.
The test for simulating the multi-material shock is contained in the second and third numerical example, and these two examples compare the numerical results by the energy-splitting schemes (ES-Godunov, ES-GRP) and other common schemes (Is-Godunov, UPV-Godunov).
The last two examples are about the interaction between shocks and material interfaces in two dimensions.
Through the comparison with the corresponding physical experimental results, the numerical results show that the current schemes for the novel reduced model perform well for two-dimensional cases with very sharp interfaces.  

\subsection{The positivity of volume fractions}\label{Sec:Why_ex}

This is an inward two-fluid compression problem, for which the initial discontinuity  at $x=0.12$ separates air with $\gamma_a=1.4$  in the left from wolfram with $\gamma_b=3.0$ in the right. 
The initial data in the entire computational domain $[0,0.15]$,  composed of $250$ cells, are given as
\begin{equation*}
\setlength\arraycolsep{0pt}
\begin{array}{lclclclr}
(\rho,u,p,\phi_a)=  (&0.00129&,&\, 0&,&\, 1.01325&,1),  &~ x<0.12,\\
(\rho,u,p,\phi_b)=  (&19.237&, &-200&, &1.01325&,1),&~ x> 0.12.
\end{array}
\end{equation*}
The left boundary is a solid wall and the right boundary has an inflow condition. This problem has exceedingly huge density ratio and velocity gradient.
We use \textit{NO-KE} to represent no kinetic energy exchange term in the scheme, and list the numerical results of $z_a$ in the $199$-th cell at time steps from $1$ to $5$ in Table \ref{why_KE}.
\begin{table}[htp]
\centering
\caption{\label{why_KE}The interfacial volume fraction $z_a$  at advancing time steps solved by different schemes.}
\begin{tabular}{crrrrr}
\hline
Scheme & \multicolumn{1}{c} \mbox{Step} 1 & \multicolumn{1}{c} \mbox{Step} 2 & \multicolumn{1}{c} \mbox{Step} 3 & \multicolumn{1}{c} \mbox{Step} 4 & \multicolumn{1}{c} \mbox{Step} 5 \\
\hline
ES-Godunov(NO-KE)& -0.09652& -0.07664& -0.05920& -0.04542& -0.03496\\
ES-GRP(NO-KE)& -0.09652& -0.07707& -0.05412& -0.03272& -0.02203
\\
ES-Godunov& 0.96143& 0.90025& 0.80975& 0.68928& 0.54981
\\
ES-GRP& 0.96143& 0.88862& 0.71710& 0.45799& 0.19412\\
\hline
\end{tabular}
\end{table} 
The numerical results show that without the process of kinetic energy exchange, the volume fraction of air at the interface  becomes negative value, which immediately ruins the numerical simulation. It shows the necessity of the kinetic energy exchange terms in the current  methods.

\subsection{Two-fluid shock-tube problem}\label{sec:sod}

This is a two-fluid shock-tube problem in  \cite{abgrall_how_1994}. The discontinuity initially at $x=0.3$ separates air with $\gamma_a=1.4,C_{v,a}=0.72$  in the left from helium with $\gamma_b=1.67,C_{v,b}=3.11$ in the right. Then the initial data in the entire computational domain $[0,1]$,  composed of $100$ cells,  are given by
\begin{equation*}
\setlength\arraycolsep{0pt}
\begin{array}{lclll}
(\rho,u,p,\phi_a) = (&1&,0,25,1), &~ x<0.3,\\
(\rho,u,p,\phi_b) = (&0.01&,0,20,1), &~ x > 0.3.
\end{array}
\end{equation*}
The exact solution of the shock-tube problem consists of a left-propagating rarefaction wave, a contact discontinuity moving at the speed of $0.83$, and a right-propagating shock wave at the speed of $58.35$. We compare the  solutions computed by  different schemes at time $t=0.008$.
\begin{figure}[htb]
\centering
\begin{minipage}[t]{0.49\linewidth}
\centering
\includegraphics[width=\textwidth]{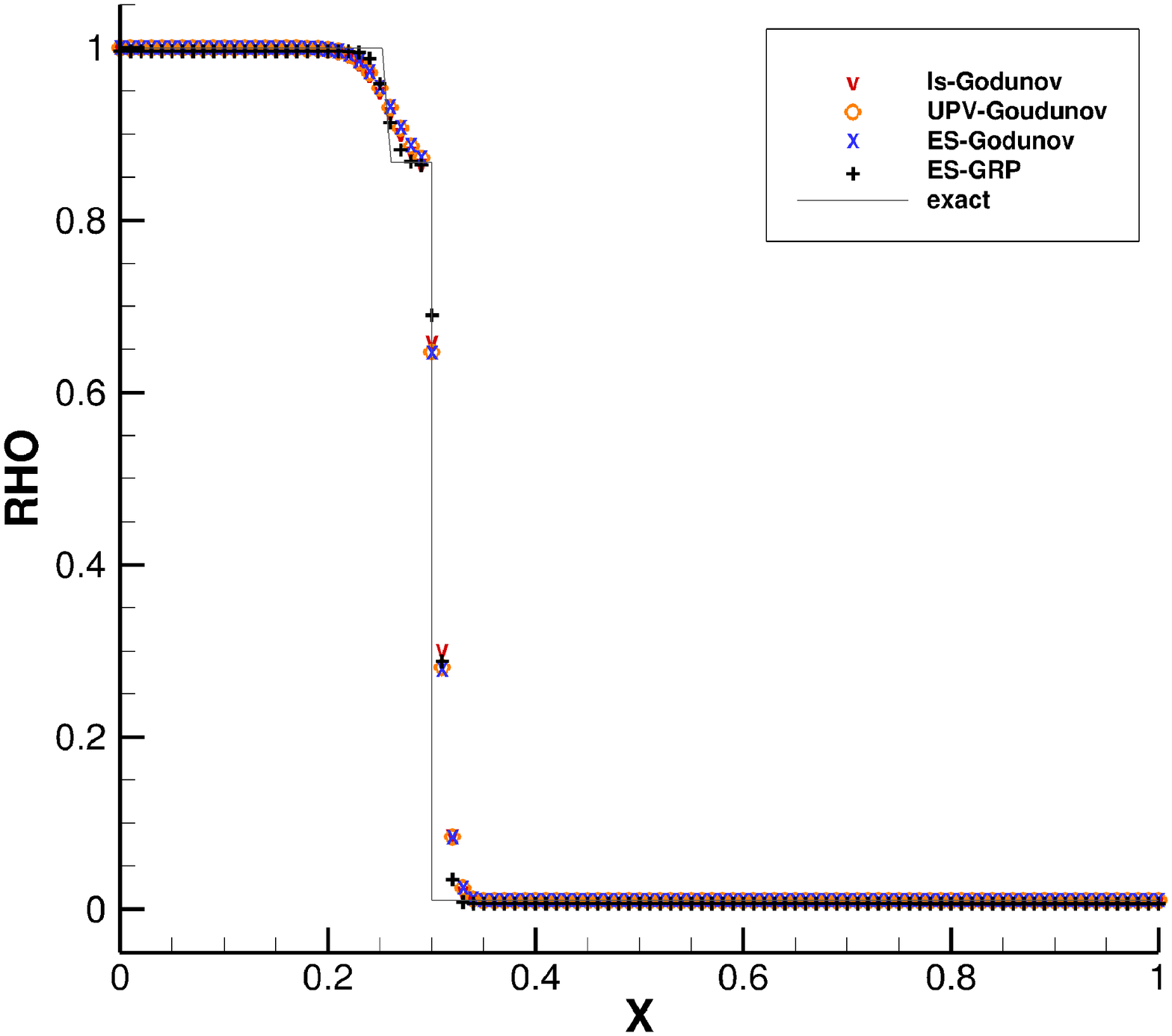}
(a)density
\end{minipage}
\hfill
\begin{minipage}[t]{0.49\linewidth}
\centering
\includegraphics[width=\textwidth]{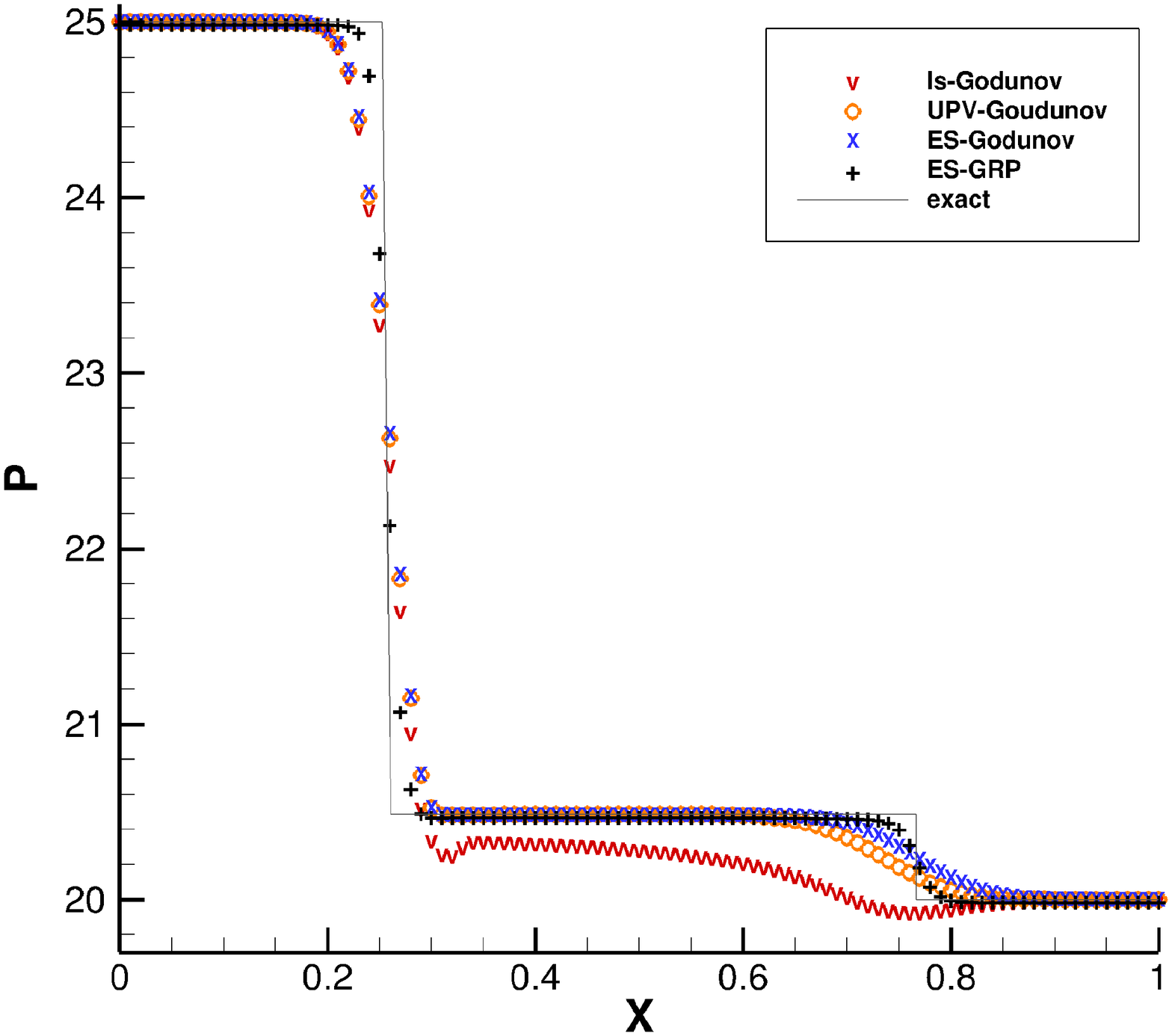}
(b)pressure
\end{minipage}
\begin{minipage}[t]{0.49\linewidth}
\centering
\includegraphics[width=\textwidth]{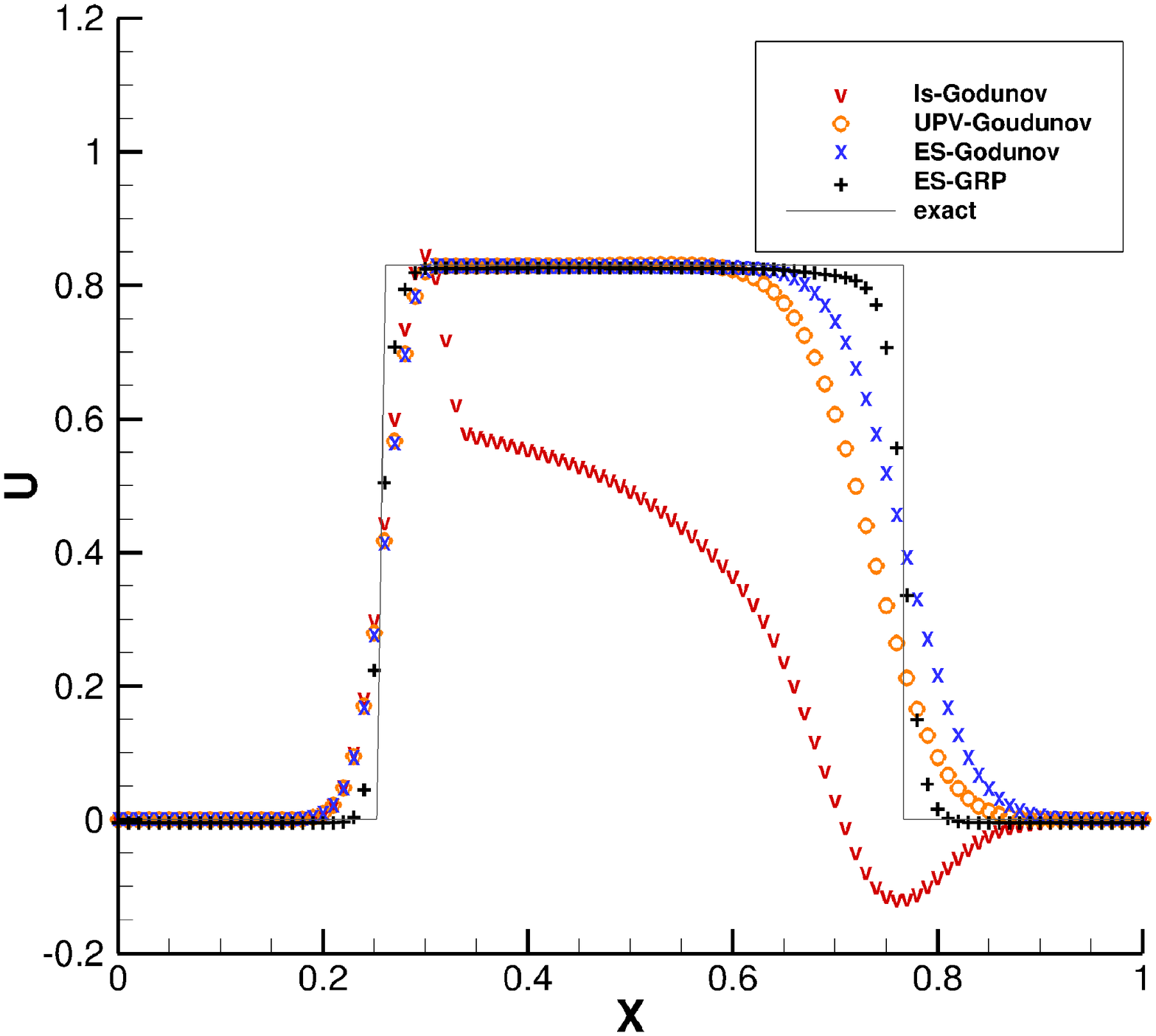}
(c)velocity
\end{minipage}
\hfill
\begin{minipage}[t]{0.49\linewidth}
\centering
\includegraphics[width=\textwidth]{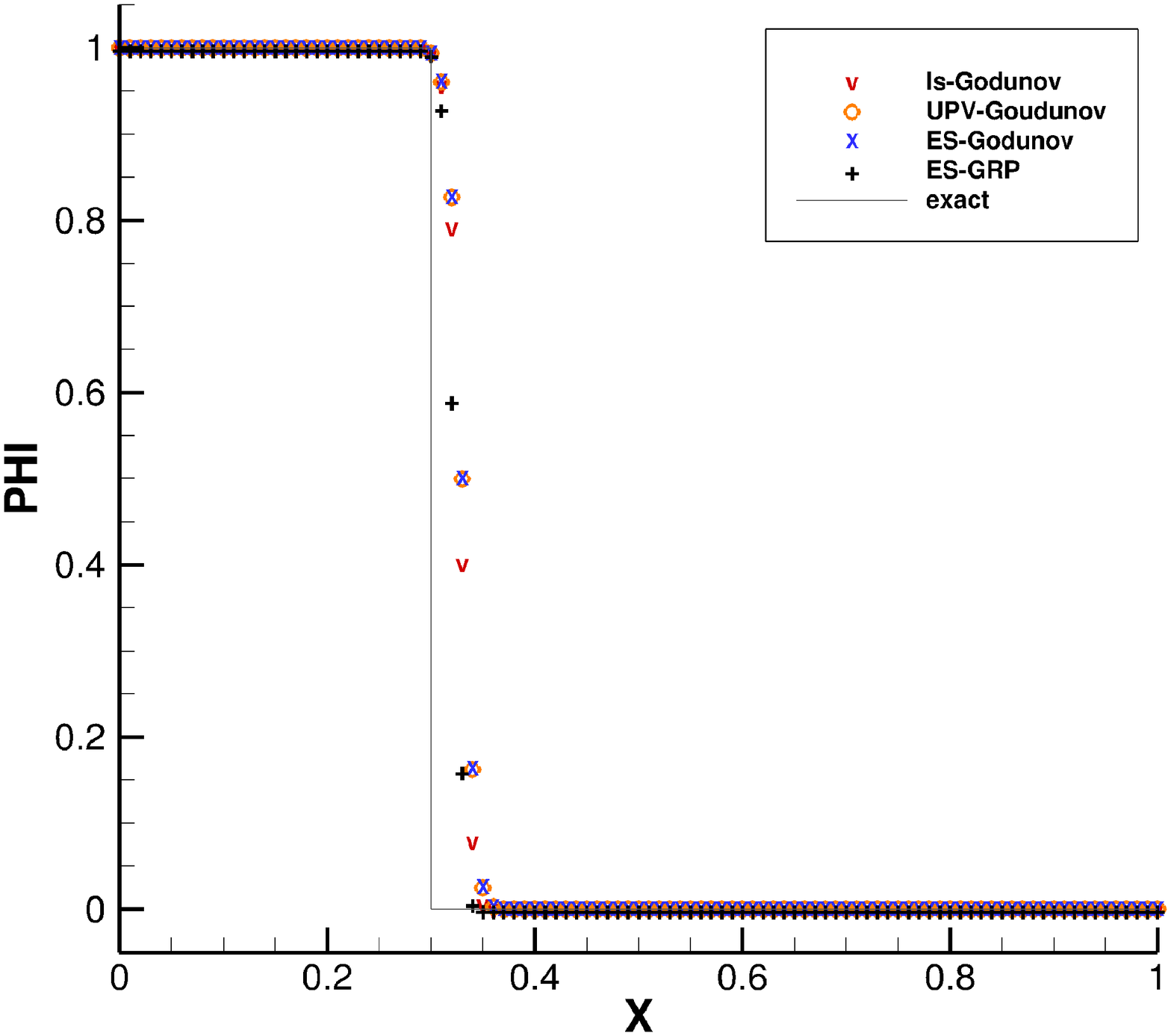}
(d)mass fraction of fluid $a$
\end{minipage}
\caption{\label{Sod}Results of the two-fluid Sod problem at $t=0.008$}
\end{figure} 
The numerical solutions computed by Is-Godunov, UVP-Godunov, ES-Godunov and ES-GRP are shown in Fig.~\ref{Sod}, in which the solid gray curves are the exact solution.
The results solved by the current energy-splitting schemes are  much closer to the exact solution than that by Is-Godunov and UPV-Godunov, which shows the good performance of the current schemes in simulating multi-material shocks.

\subsection{Shock-interface interaction}\label{sec:int_err}

\begin{figure*}[htb]
\begin{minipage}[t]{0.49\linewidth}
\centering
\includegraphics[width=\textwidth]{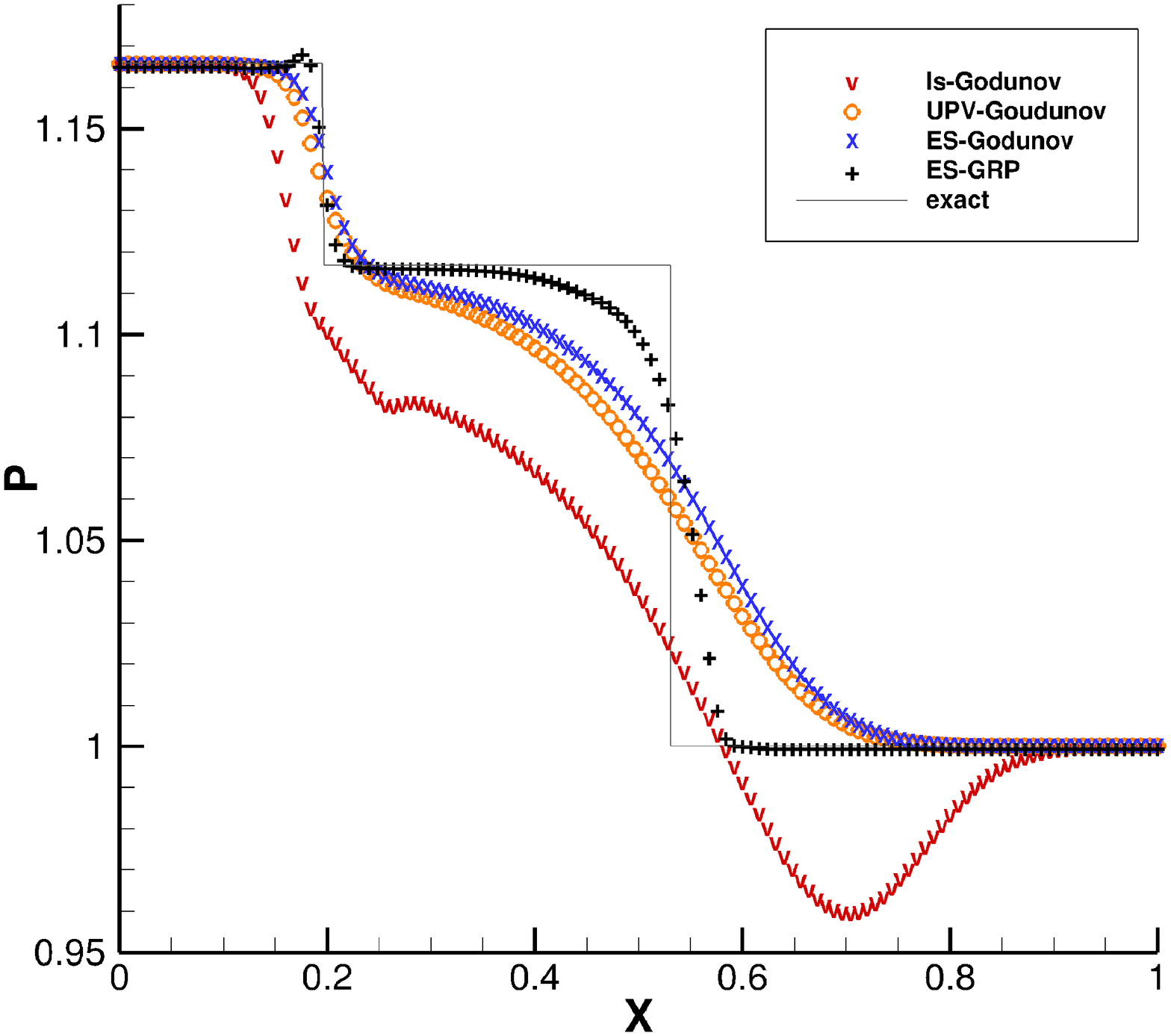}
(a)pressure
\end{minipage}
\hfill
\begin{minipage}[t]{0.49\linewidth}
\centering
\includegraphics[width=\textwidth]{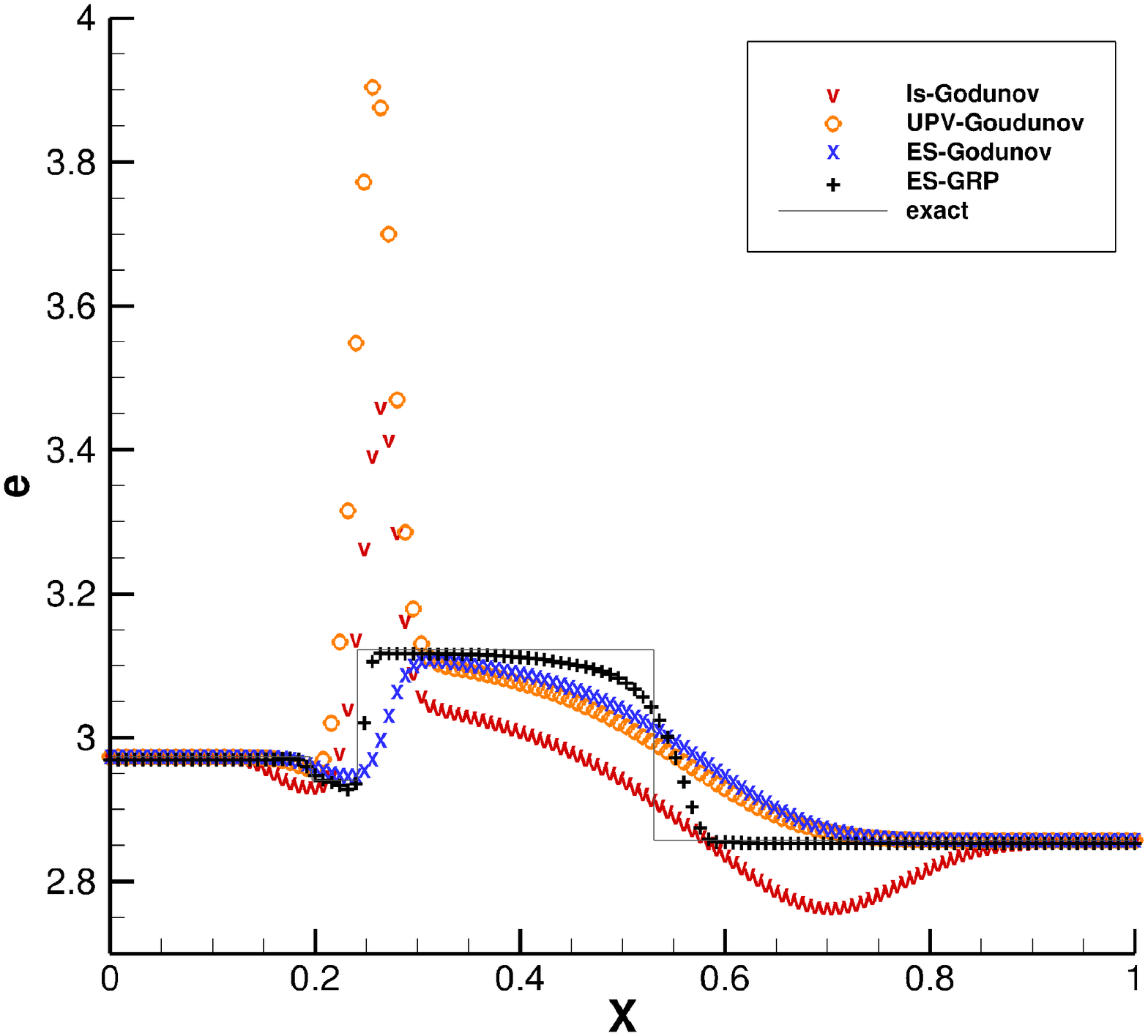}
(b)specific internal energy
\end{minipage}
\caption{\label{SIM}Results of the shock-interface interaction problem at $t=0.07$.}
\end{figure*}

This is a shock-interface interaction problem.  The interface initially at $x=0.2$ separates material $a$  with $\gamma_a=1.35,C_{v,a}=2.4$ in the left from material $b$ with $\gamma_b=5.0,C_{v,b}=1.5$ in the right. These two materials correspond to high explosive products in the left and a confining material in the right \cite{banks_high-resolution_2007}. The interface and a shock wave with the shock Mach number $M_s=1.5$ initially at $x=0.16$ propagate to the right at the speed of $0.5$ and $1.74$, respectively. Then the initial data in the computational domain $[0,1]$,  composed of $125$ cells,  are given by
\begin{equation*}
\setlength\arraycolsep{0pt}
\begin{array}{lclclclr}
(\rho,u,p,\phi_a)=  (&1.1201&,&\, 0.6333&,&\, 1.1657&,1),  &~ x<0.16,\\
(\rho,u,p,\phi_a)=  (&1&, &0.5&, &1&,1),  &~ 0.16< x<0.2,\\
(\rho,u,p,\phi_b)=  (&0.0875&, &0.5&, &1&,1),&~ x> 0.2.
\end{array}
\end{equation*}
At  time $t=0.0322$, the interface is impacted by the shock wave. The resulting wave pattern after the interaction consists of a reflected rarefaction wave, an interface at the speed of $0.67$, and a transmitted shock at the speed of $8.32$. We compare the profiles of pressure and internal energy by using different methods at $t=0.07$ in Fig.~\ref{SIM}.
Serious error of the internal energy occur at the interface in Is-Godunov and  UVP-Godunov solutions. 
In contrast, the current method can produce much better results.

\subsection{Shock-bubble interactions}

\begin{figure}[ht]
\centering
\includegraphics[width=0.45\textwidth]{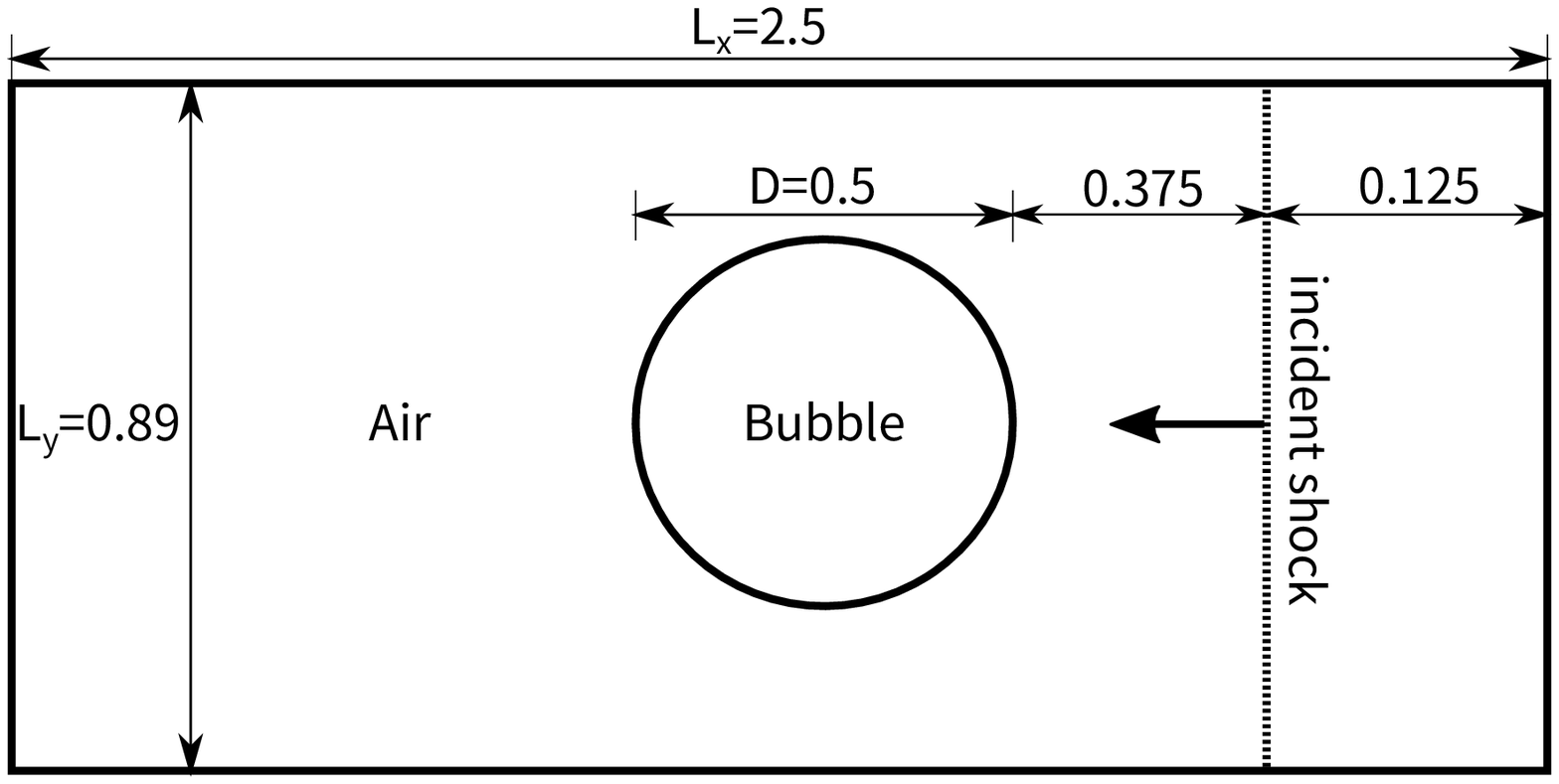}
\caption{\label{bubble}Diagram of the shock-bubble interaction problem}
\end{figure}

This example is about the interaction problem of a planar shock wave with a cylindrical gas bubble. This problem is motivated by the experiments in \cite{haas_interaction_1987}.
In the experiments, a weak shock with the shock Mach number $M_s=1.22$ propagates from atmospheric air into a stationary cylindrical bubble filled with lighter helium or heavier Refrigerant 22 (R22). The computational domain $[0,2.5]\times[0,0.89]$ composes of $2500\times 890$ square cells and the position of initial discontinuity is set in Fig.~\ref{bubble}.
The upper and lower boundaries are solid wall boundaries, whereas the left and right boundaries are non-reflective. The air outside and the gas inside the bubble are assumed initially to be in the temperature and pressure equilibrium.
\begin{table}[ht]
\centering
\caption{\label{parameters}Some parameters for the shock-bubble interaction problems in front of the shock wave}
\begin{tabular}{cccc}
\hline
Gas & \multicolumn{1}{c}{\mbox{Air}} &  \multicolumn{1}{c}{\mbox{Helium}+28\%\mbox{Air}} & \multicolumn{1}{c}{\mbox{R22}}\\
\hline
$\gamma$   &1.40   &1.648 &1.249\\
$C_v$   &0.72   &2.44 &0.365\\
$\rho$   &1   &0.182 &3.169\\
$p$ &1&1&1\\
$u$ &0&0&0\\
\hline
\end{tabular}
\end{table}
For the helium bubble case, the gas in the bubble is assumed as a helium-air mixture where the mass fraction of air is $28\%$, which is explained in \cite{haas_interaction_1987}. These materials are regarded as ideal gases, and the corresponding fluid state and parameters taken from \cite{quirk_dynamics_1996} are presented in Table \ref{parameters}.

\vspace{0.2cm}
 
\begin{figure}[htb]
\centering
\begin{minipage}[t]{0.49\textwidth}
\begin{minipage}[t]{\textwidth}
(a)
\hfill
\includegraphics[width=0.9\textwidth]{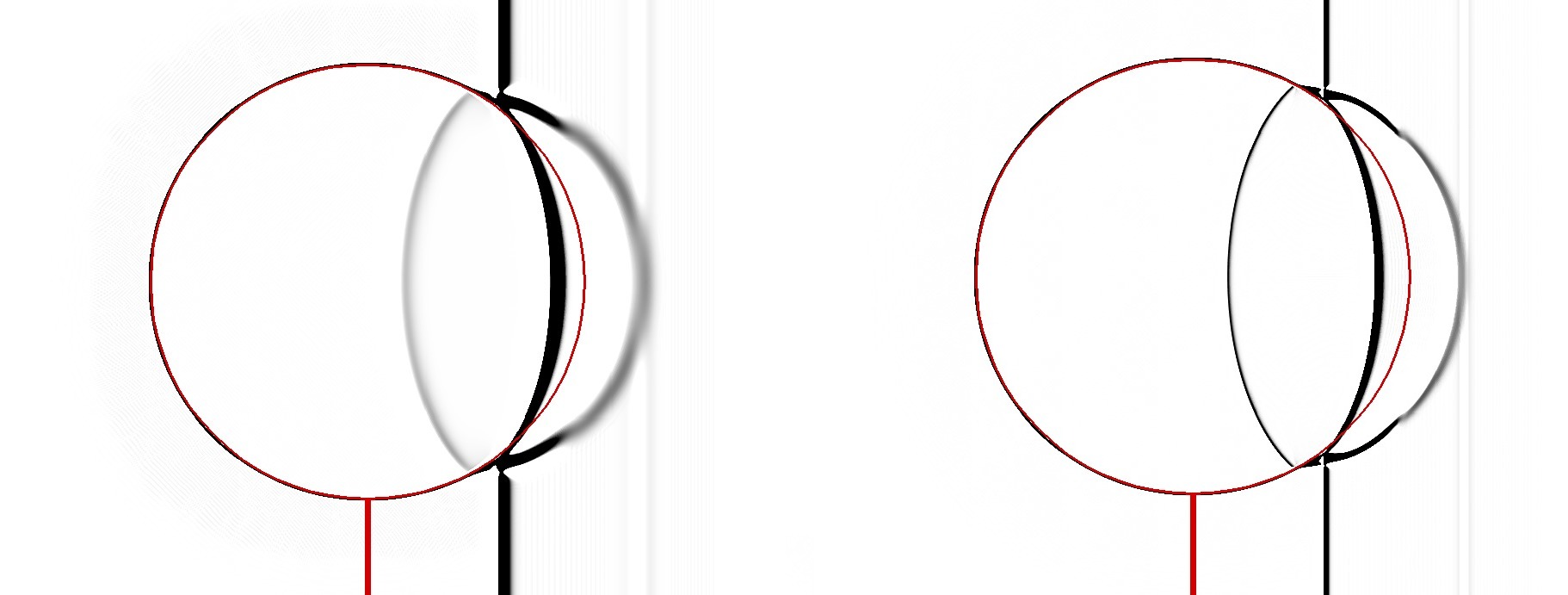}
\end{minipage}
\vfill
\begin{minipage}[t]{\textwidth}
(b)
\hfill
\includegraphics[width=0.9\textwidth]{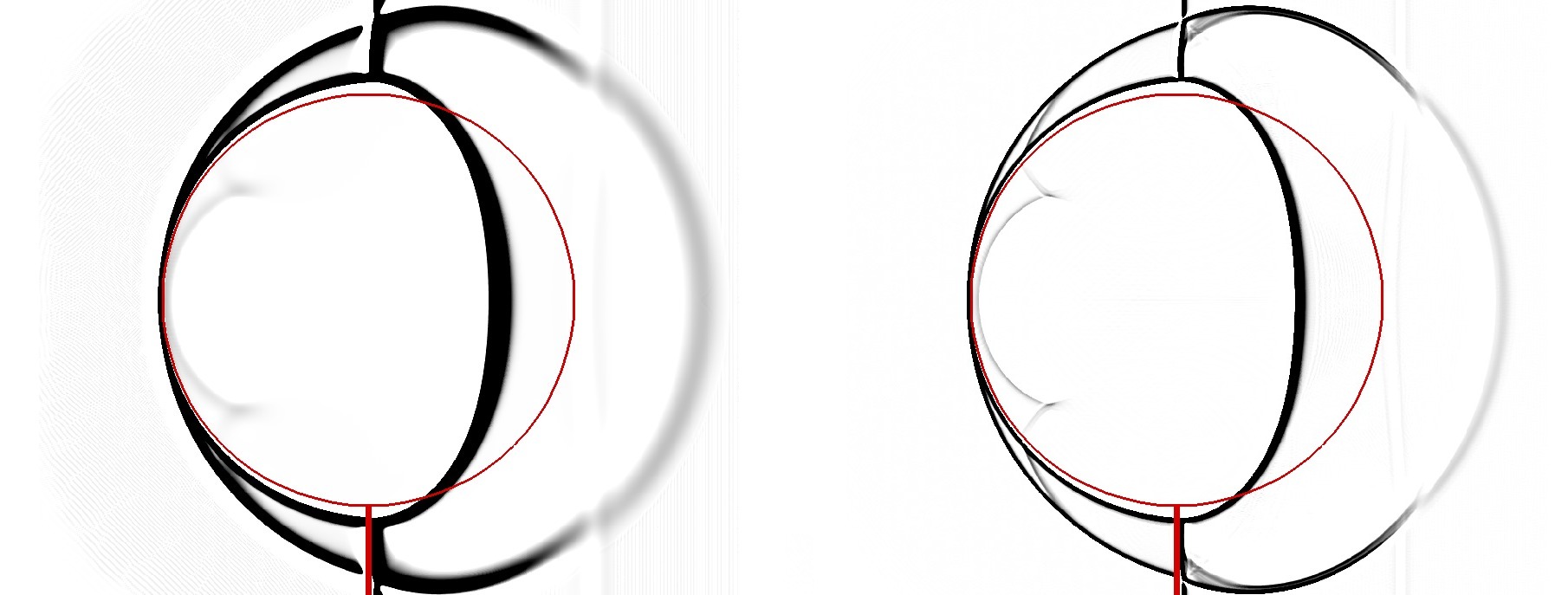}
\end{minipage}
\vfill
\begin{minipage}[t]{\textwidth}
(c)
\hfill
\includegraphics[width=0.9\textwidth]{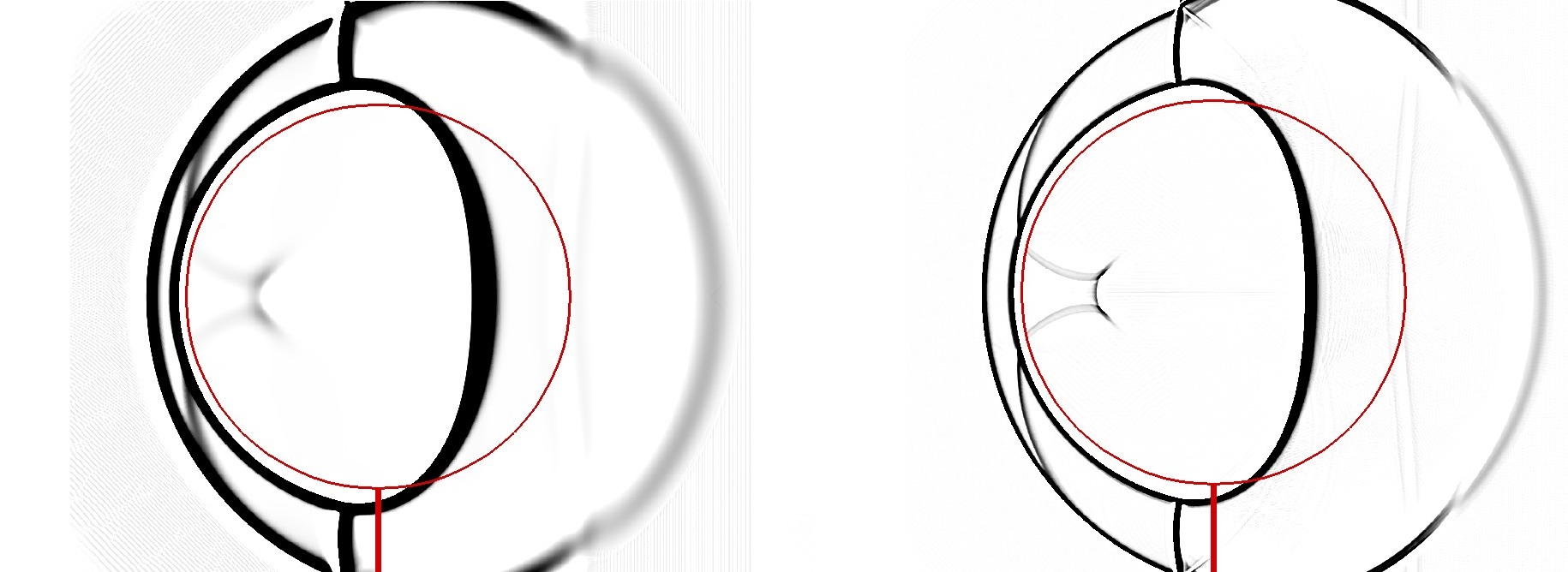}
\end{minipage}
\vfill
\begin{minipage}[t]{\textwidth}
(d)
\hfill
\includegraphics[width=0.9\textwidth]{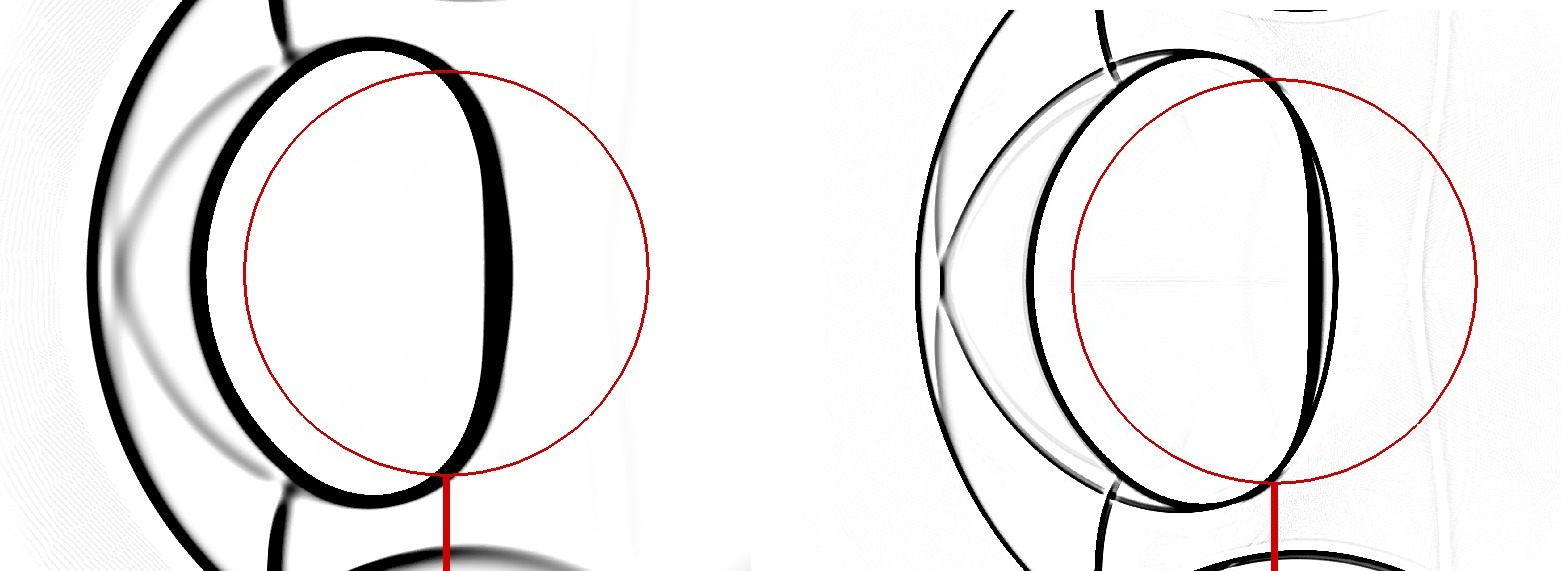}
\end{minipage}
\vfill
\begin{minipage}[t]{\textwidth}
(e)
\hfill
\includegraphics[width=0.9\textwidth]{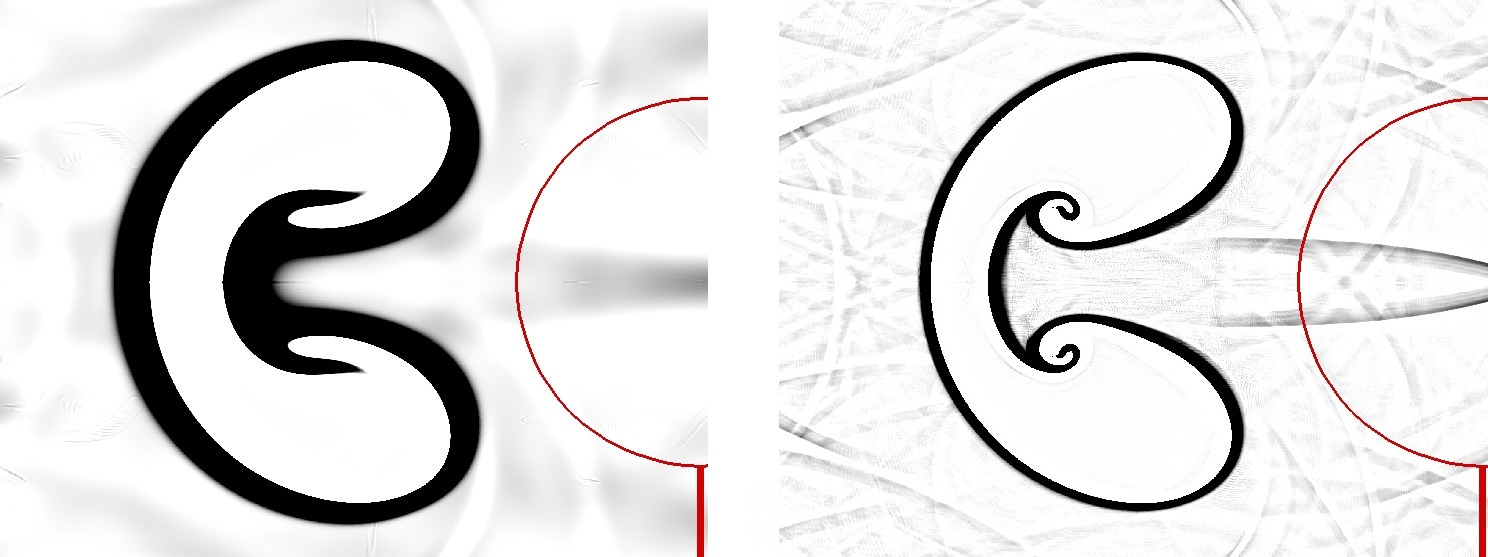}
\end{minipage}
\vfill
\begin{minipage}[t]{\textwidth}
(f)
\hfill
\includegraphics[width=0.9\textwidth]{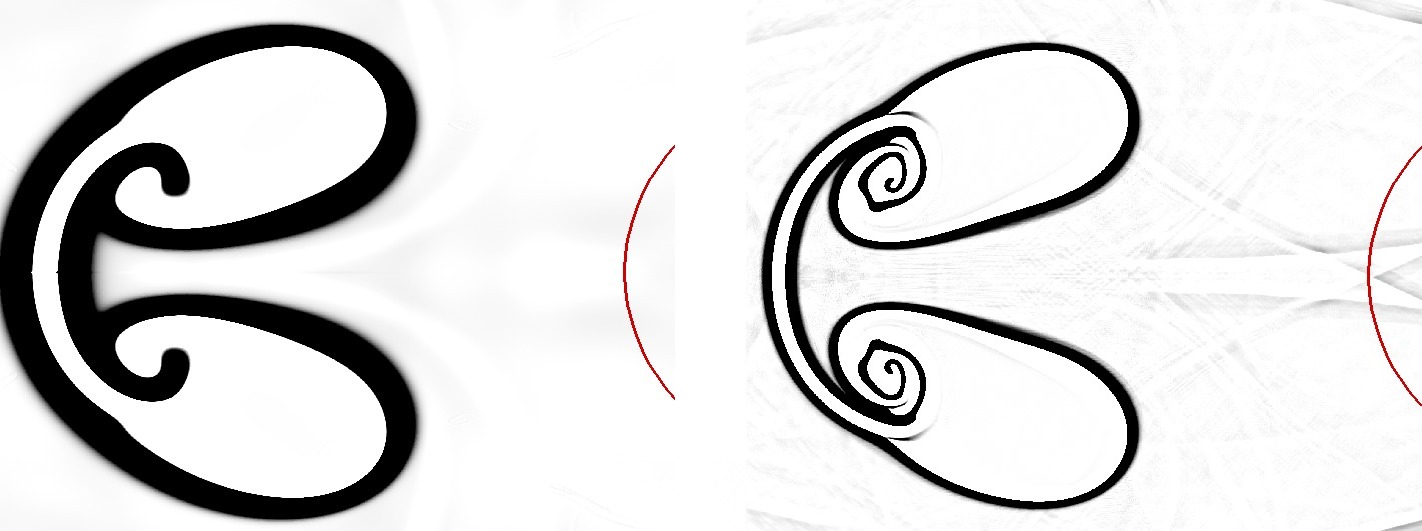}
\end{minipage}
\end{minipage}
\hfill
\begin{minipage}[t]{0.49\textwidth}
\begin{minipage}[t]{\textwidth}
(a)
\hfill
\includegraphics[width=0.9\textwidth]{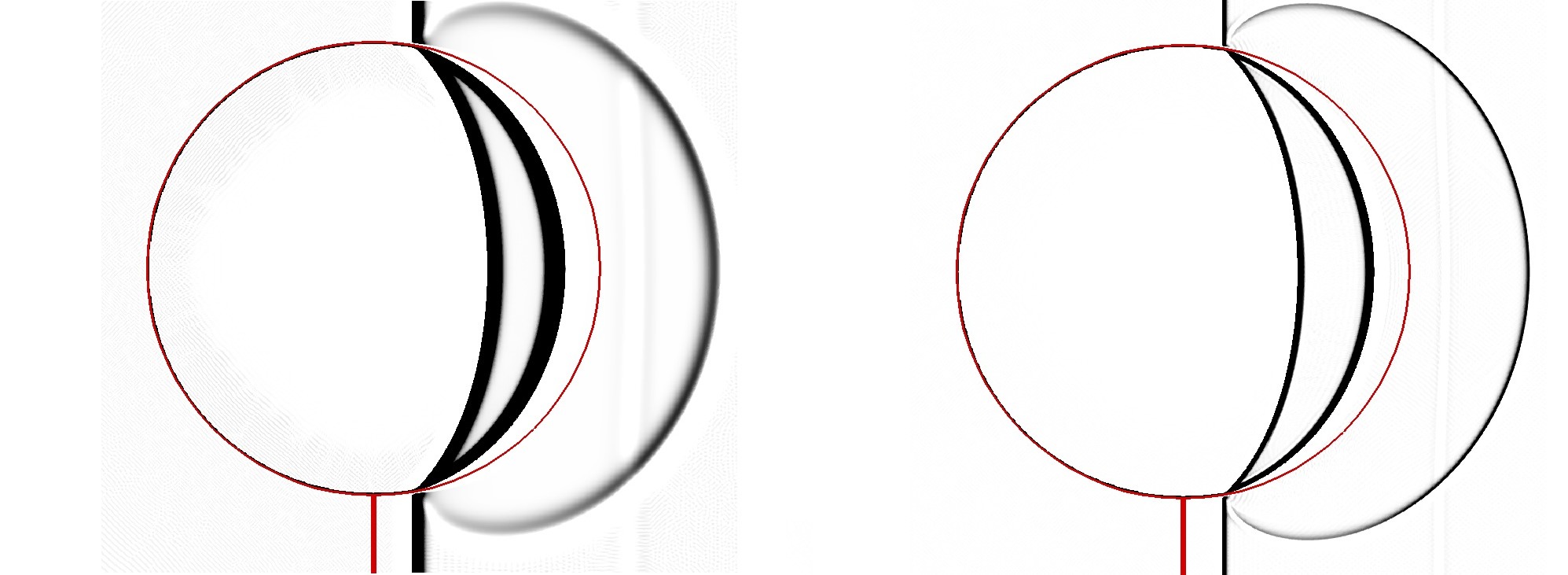}
\end{minipage}
\vfill
\begin{minipage}[t]{\textwidth}
(b)
\hfill
\includegraphics[width=0.9\textwidth]{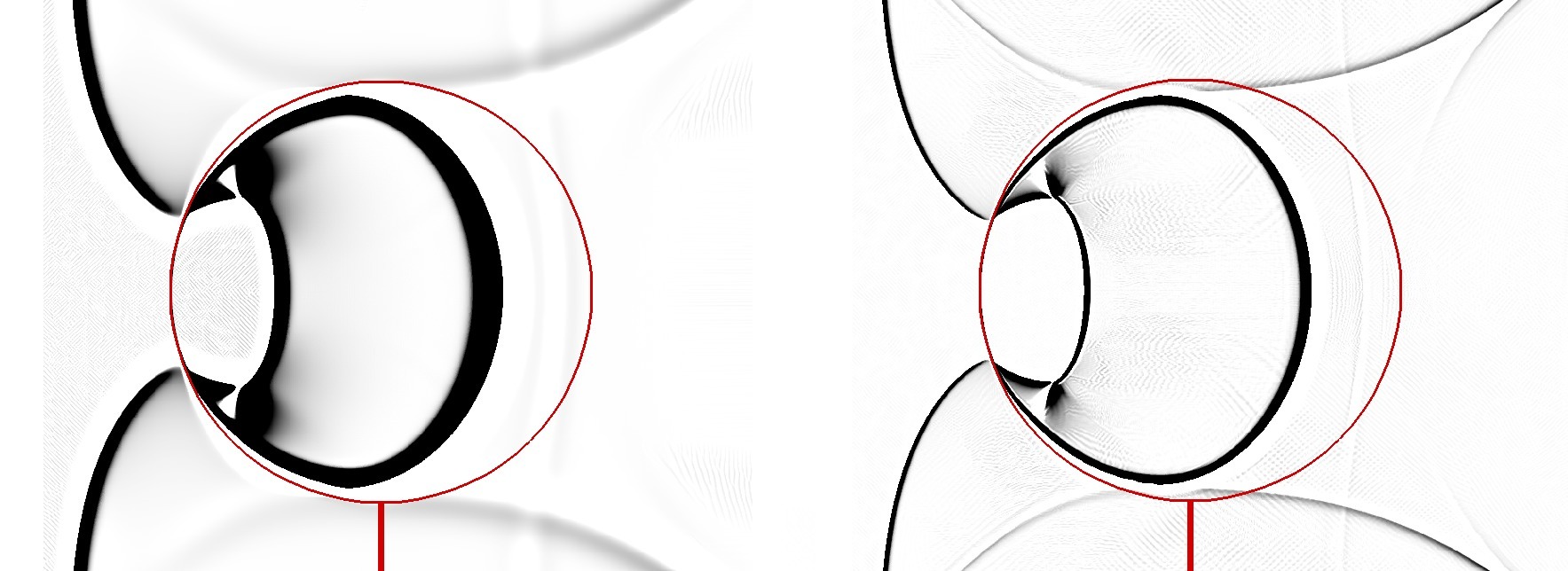}
\end{minipage}
\vfill
\begin{minipage}[t]{\textwidth}
(c)
\hfill
\includegraphics[width=0.9\textwidth]{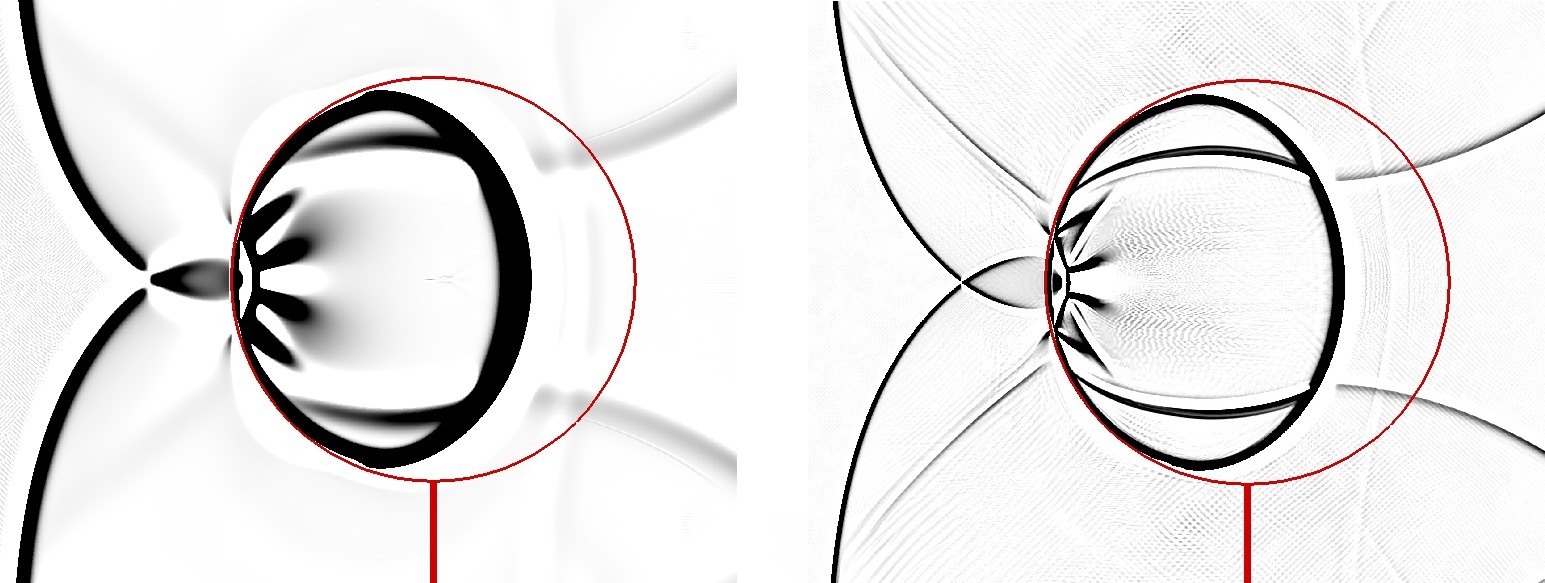}
\end{minipage}
\vfill
\begin{minipage}[t]{\textwidth}
(d)
\hfill
\includegraphics[width=0.9\textwidth]{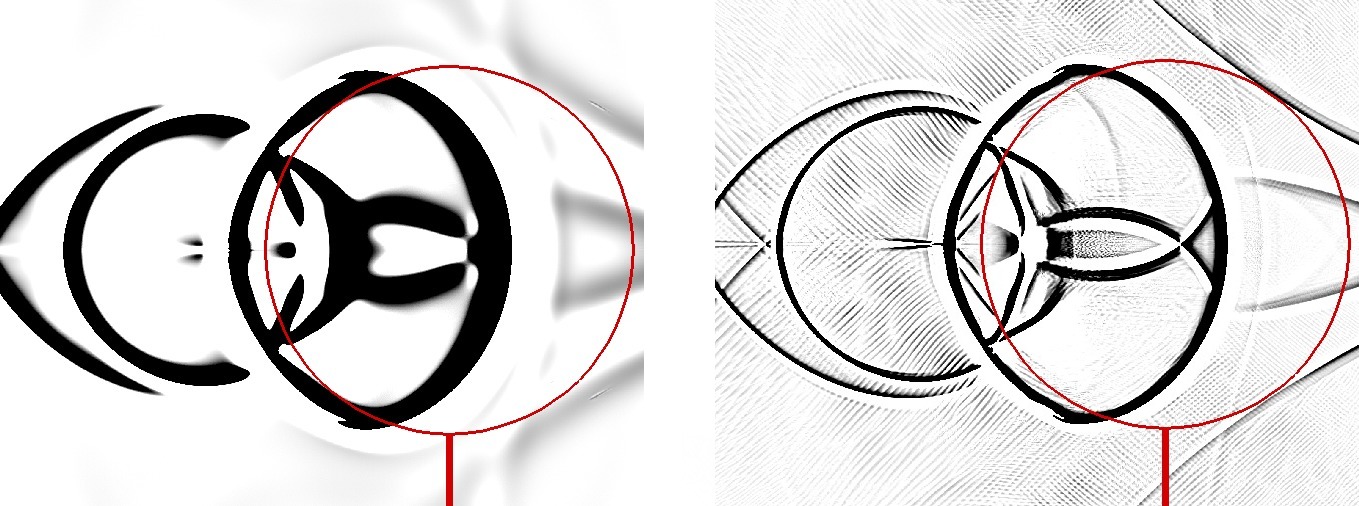}
\end{minipage}
\vfill
\begin{minipage}[t]{\textwidth}
(e)
\hfill
\includegraphics[width=0.9\textwidth]{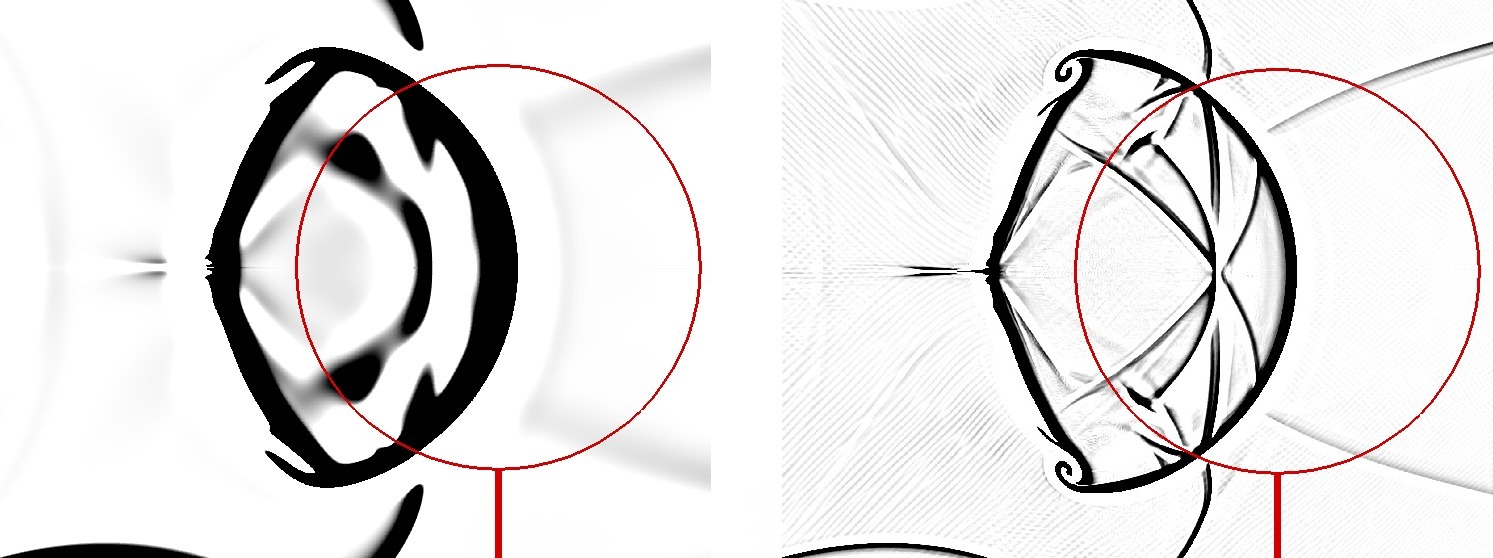}
\end{minipage}
\vfill
\begin{minipage}[t]{\textwidth}
(f)
\hfill
\includegraphics[width=0.9\textwidth]{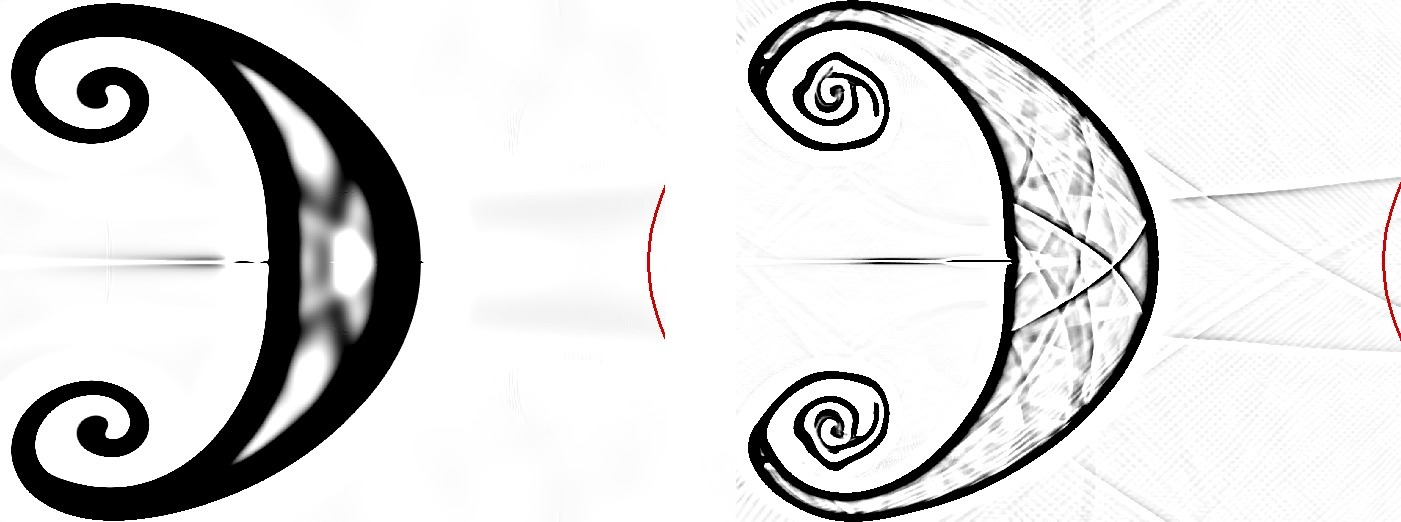}
\end{minipage}
\end{minipage}
\caption[small]{Numerical shadow-graph images of the shock bubble interaction with $M_s=1.22$.
On the left is the results of helium bubble obtained by ES-Godunov (column 1) and ES-GRP (column 2) at experimental times ($\mu$s): (a)$32$, (b)$62$, (c)$72$, (d)$102$, (e)$427$ and (f)$674$;
on the right is the results of R22 bubble obtained by ES-Godunov (column 3) and ES-GRP  (column 4) at experimental times ($\mu$s): (a)$55$, (b)$135$, (c)$187$, (d)$247$, (e)$342$ and (f)$1020$. The corresponding experimental shadow-photographs can be found in  \cite[Figures 7 and 11]{haas_interaction_1987}.}\label{Helium}
\end{figure}

Fig.~\ref{Helium} compares the numerical shadow-graph images of the shock-helium bubble interaction problem and the shock-R22 bubble interaction problem by ES-Godunov and ES-GRP, corresponding to the experiments at different  times in \cite{haas_interaction_1987}. In order to better compare the results, the initial interface (red curves) is  added to the numerical shadow-graph images.
Since the sound speed of helium inside the bubble is much greater than the sound speed of the air outside, the helium bubble acts as a divergent lens for the incident shock.
In contrast, as the sound speed of R22 inside the bubble is much lower than that of the air outside, the R22 bubble acts as a convergent lens.
The numerical shadow-graph images show a very good agreement between the second-order numerical simulations and the laboratory experiments. As ES-GRP is used, the stability issue becomes weaker along the material interface than the numerical results in \cite{quirk_dynamics_1996} and much clearer discontinuity surfaces are observed than those by ES-Godunov.

\subsection{Two-fluid Richtmyer-Meshkov instability}

This test problem concerns the multi-fluid Richtmyer-Meshkov instability (RMI) phenomenon in two dimensions.
The RMI phenomenon was originally theoretically predicted by Richtmyer, and subsequently observed in experiments by Meshkov.
Here, we simulate a simple test example of a plane shock hitting a sinusoidal perturbed interface, separating SF$_6$ and air.
\begin{figure}[htb]
\begin{center}
\includegraphics[width=0.5\textwidth]{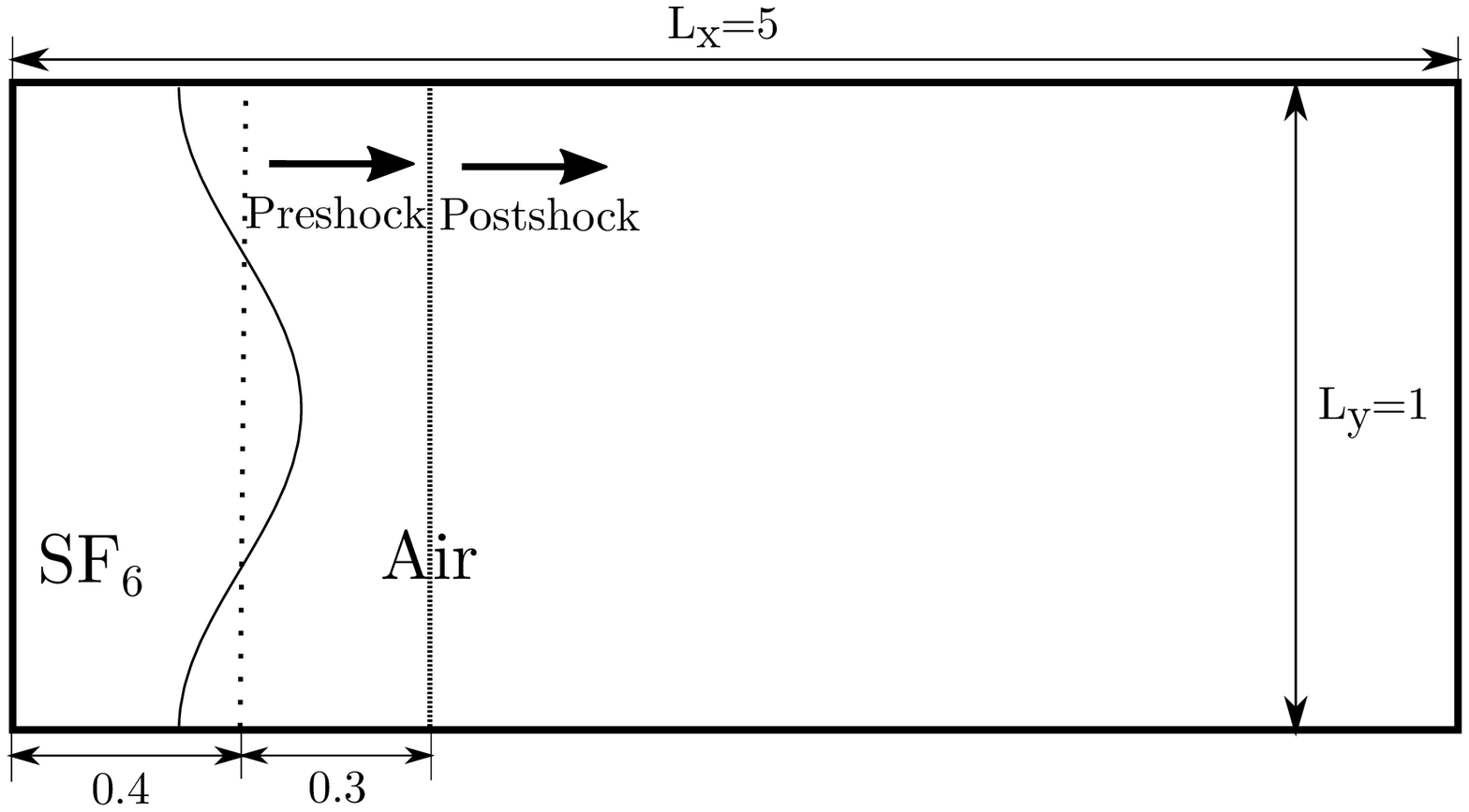}
\caption{Diagram of the two-fluid RMI test problem}\label{fig:RMI-init}
\end{center}
\end{figure}
A schematic diagram of the initial flow configuration is shown in Fig.~\ref{fig:RMI-init}, and the initial configuration of the perturbed interface is generated with \cite{nonomura_numerical_2012}
\begin{equation*}
x_d=0.4 + 0.1 \sin(2\pi (y + 0.25)), -0.5 < y < 0.5, 
\end{equation*}
which separates the SF$_6$ with $\gamma_a = 1.094$ in the left from the air with $\gamma_b =1.4$ in the right.
The initial conditions are shown in Table \ref{tab:parameters-RMI} from two cases:
\begin{itemize}
\item[(\romannumeral1).]\  The initial density of SF$_6$ is $\rho_0=5.04$. The computational domain  is $[0,16]\times [-0.5, 0.5]$ composed of $2048\times 128$ square cells.

\item[(\romannumeral2).]  \ The initial density of SF$_6$ is $\rho_0=0.0005$. The computational domain   is $[0,5]\times [-0.5, 0.5]$ composed of $640\times 128$ square cells.
\end{itemize}
The upper and lower boundaries satisfy a symmetric boundary condition, whereas the left and right boundaries are free boundaries.
The initial conditions are put in Table \ref{tab:parameters-RMI}, including a material interface with a large density difference.
\begin{table}[ht]
\centering
\caption{Initial data of the two-fluid RMI test problems}\label{tab:parameters-RMI}
\begin{tabular}{cccccc}
\hline
Position & Gas & $\rho$ & $p$ & $u$ & $v$\\
\hline
$x<x_d$ & SF$_6$ & $\rho_0$& $1.24$& $0.7143$ & 0\\
$x_d< x<0.7$ & Air & $1$ & $1.24$ & $0.7143$ & 0\\
$0.7 <x< 7$ & Air & $1.4112$ & $0.8787$ & $1.1623$ & $0$\\
\hline
\end{tabular}
\end{table}
For these two cases, the development of instability is sensitive to the initial perturbation. The incident shock hits the material interface, resulting in a transmitted shock and a reflected shock. The accelerated interface forms a rolled-up spike shape.

\begin{figure}[htb]
\center
\begin{minipage}{0.49\linewidth}
\begin{center}
\includegraphics[width=\textwidth]{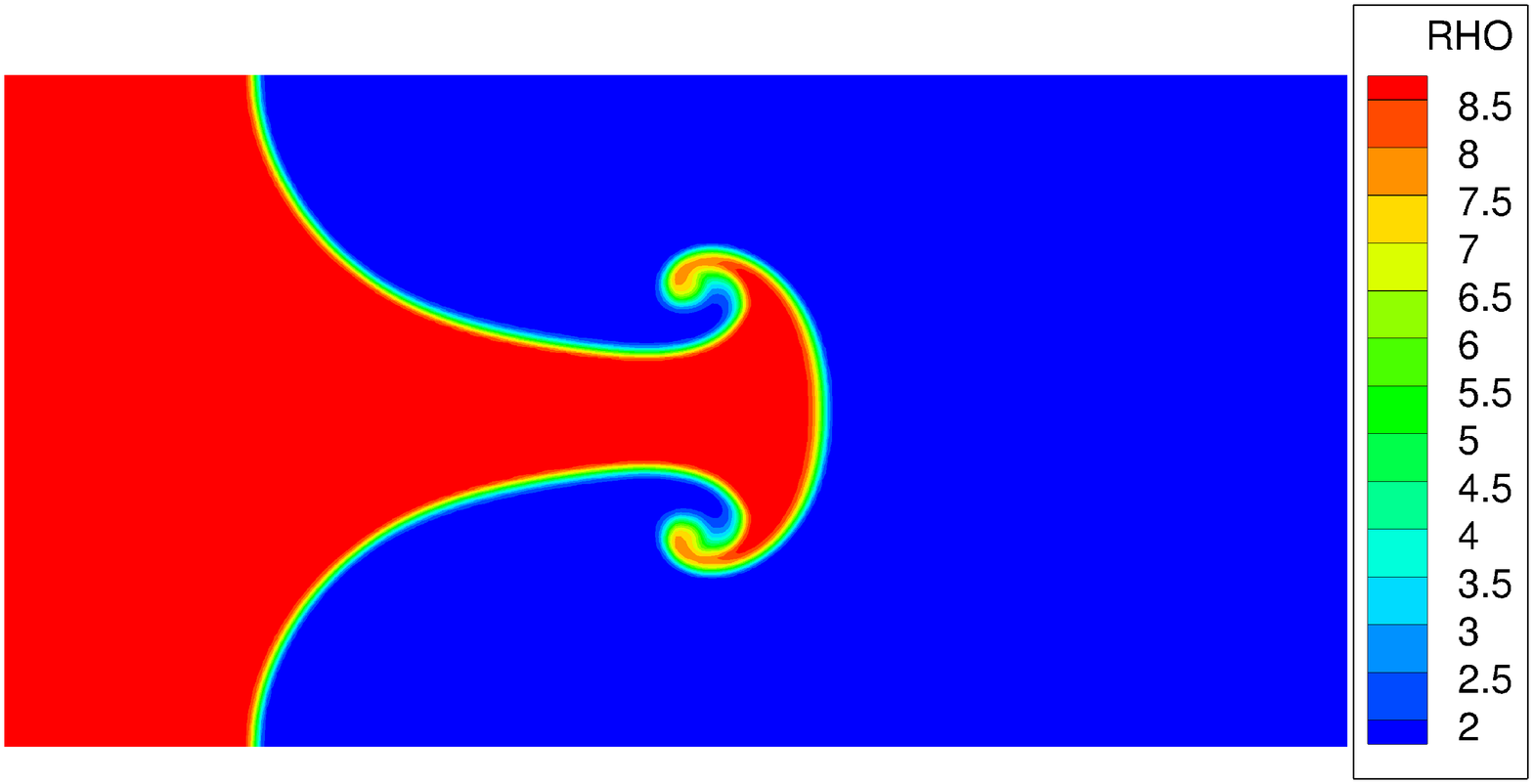}
(a) 1-D GRP solver
\end{center}
\end{minipage}
\begin{minipage}{0.49\linewidth}
\begin{center}
\includegraphics[width=\textwidth]{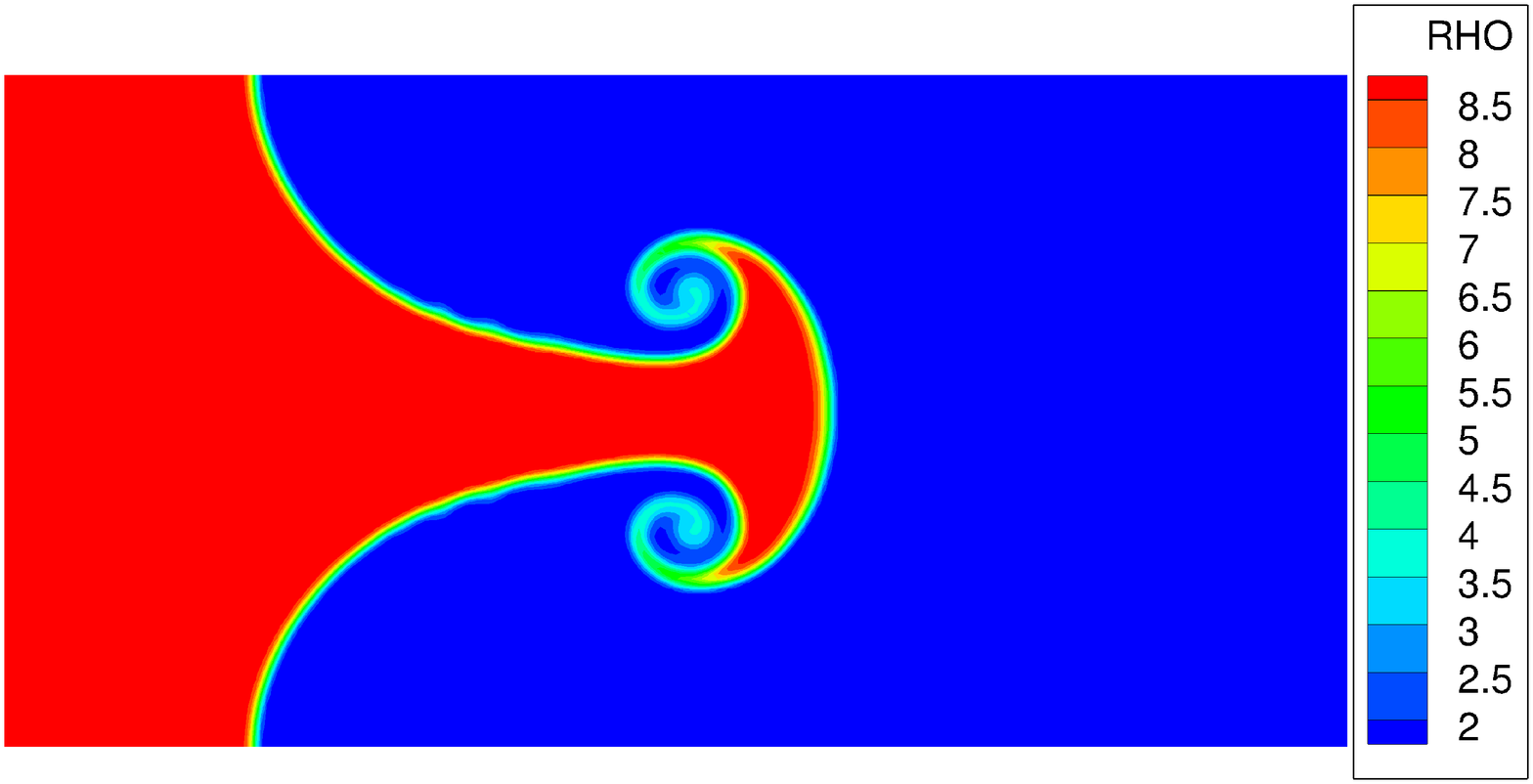}
(b) 2-D GRP solver
\end{center}
\end{minipage}
\caption{Density distribution for the two-fluid RMI problem (Case (\romannumeral1)) by ES-GRP at $t=8.25$.}\label{fig:RMI1-phi}
\end{figure}

The numerical results of  Case {(\romannumeral1)} at $t=8.25$ are displayed in Fig.~\ref{fig:RMI1-phi}.
For the same problem,
the numerical results based on conservative variables in \cite[Fig.~25]{nonomura_numerical_2012} generate non-physical oscillations at the material interface, which is not observed in the results by ES-GRP.
\begin{figure}[htb]
\center
\begin{minipage}{0.7\linewidth}
\begin{center}
\includegraphics[width=\textwidth]{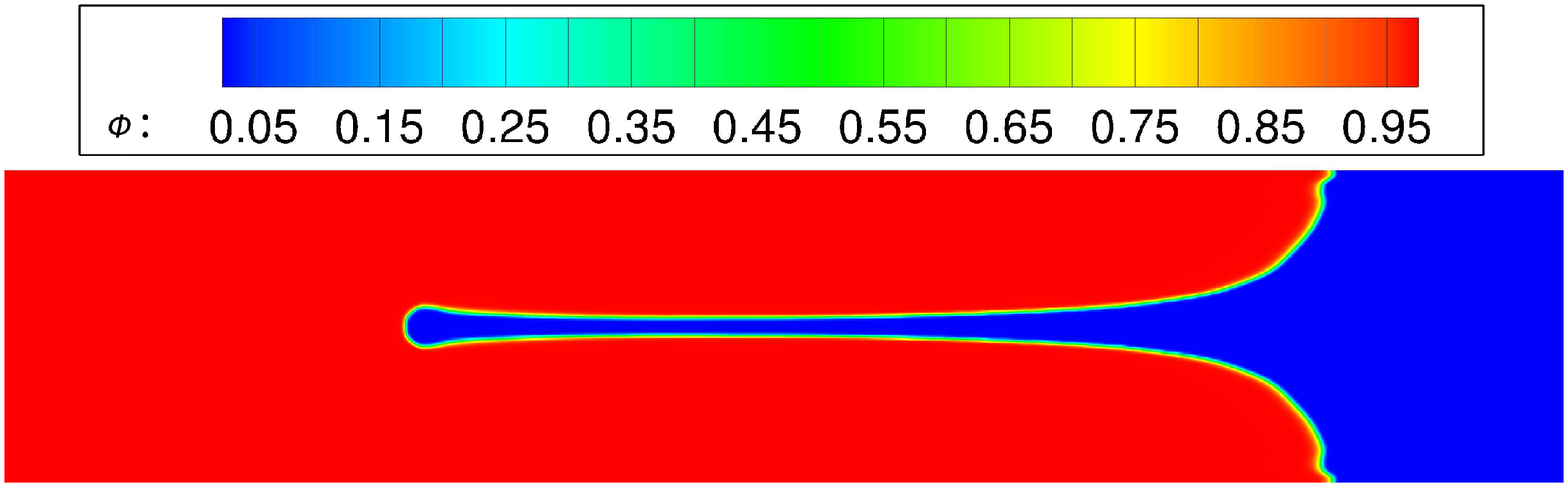}
(a) 1-D GRP solver
\end{center}
\end{minipage}
~\\
\begin{minipage}{0.7\linewidth}
\begin{center}
\includegraphics[width=\textwidth]{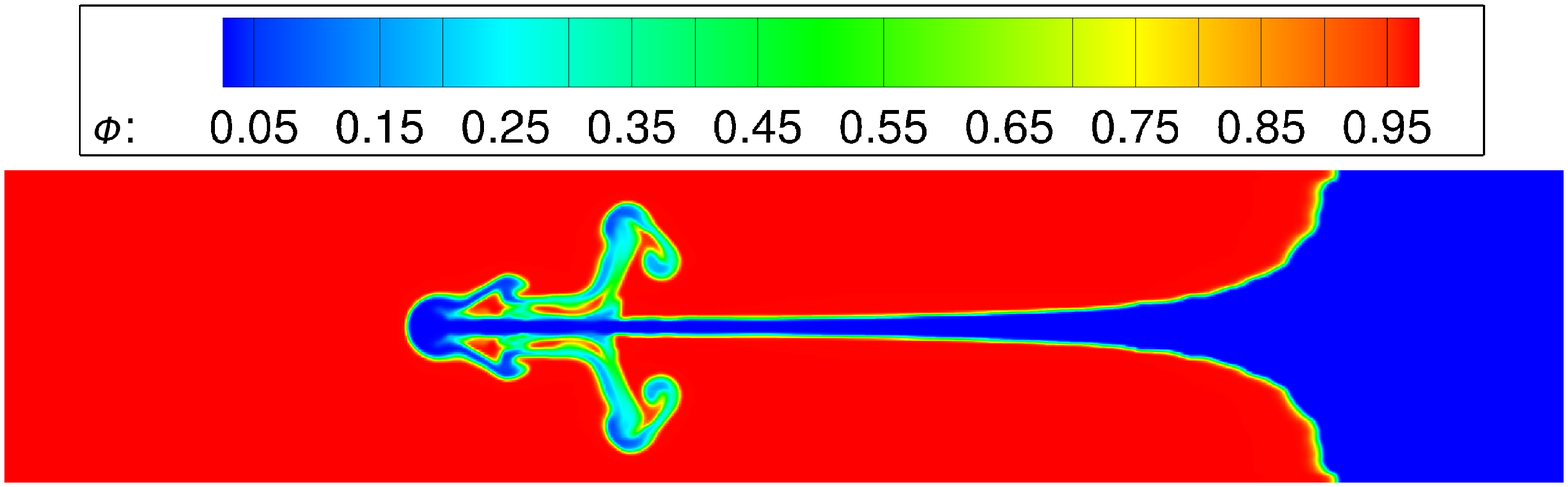}
(b) 2-D GRP solver
\end{center}
\end{minipage}
\caption{Mass fraction distribution for the two-fluid RMI problem  (case \romannumeral2) by ES-GRP at $t=6.5$. See \cite{Lei-Li-2019}.}\label{fig:RMI-phi}
\end{figure}
The numerical results at $t=6.5$ of Case ({\romannumeral2})  by ES-GRP are shown in Fig.~\ref{fig:RMI-phi}.
It is observed in these figures that  ES-GRP with the 2-D GRP solver obtains the better resolution for vortex structures compared with those  by the 1-D GRP solver.
The differences between the two solvers are described in Appendix, reflecting the ability of the 2-D GRP solver in capturing transversal effects.

\section{Conclusion}

The study of compressible multi-fluid flows is an important topic in theory, numerics and applications, and it was carried out in various ways such as physical experiments, physical modelings, numerical simulations and many others.
In this chapter, we focus on the design of numerical schemes with numerical demonstrations based on a reduced  version of the BN model.

Various reduced models based on different simplifying assumptions are presented in this chapter, and the corresponding numerical schemes are even various.
Here we show some of the most typical numerical schemes for reduced models.
The solutions by these numerical schemes are often deficient in capturing multi-material shocks or achieving real solutions of the complete BN model.
To remedy this deficiency, a novel energy-splitting scheme is designed for a novel reduced model,  based on the Godunov scheme with a second order extension by using the GRP solver.
In a sense, this scheme proposed here is compatible with the five-equation reduced model \cite{murrone_five_2005}.
However, since the novel reduced model is similar to the conservative Euler system in form, the shortcomings of non-conservative schemes can be alleviated to some extent, especially when multi-material shocks are simulated. 
In addition, the positivity preserving of volume fractions is pivotal as a numerical fluid mixing rule around  interfaces, for which the exchange of kinetic energy is considered in the current scheme so that no pressure oscillations arise near material interfaces, even though there is large difference in thermodynamic quantities.

Several benchmark problems are tested  in order to  demonstrate the validity and performance   of the current  method. The one-dimensional problems display more accurate computation of internal energy around material interfaces.
Numerical results show that the energy-splitting scheme is effective for the volume fraction positivity and the simulation of multi-material shock waves.
The two-dimensional shock-bubble interaction problems demonstrate  the performance of ES-GRP capturing  material interfaces, through the  comparison with the corresponding physical experiments. 
It is expected that this method improve the validity of the reduced forms of the BN model for simulating multi-material flows.


\section*{Appendix: The 2-D GRP solver}

Since the two-dimensional case is considered, we need to solve a so-called quasi 1-D GRP of \eqref{eq:5-eq} by setting the adjacent interface along $x = 0$, \begin{equation}\label{Q1-GRP}
\begin{array}{l}
\bm{W}_t+\operatorname{div} \bm{F}(\bm{W})=\bm{0},\\[3mm]
(z_{a})_t+\bm{u} \cdot \nabla z_{a}=\omega\operatorname{div} \bm{u},\\[3mm]
\bm{V}(x,y,0)=\left\{
\begin{array}{ll}
\bm{V}_-(x,\tilde y), \ \ \ \ & x<0, \\
\bm{V}_+(x,\tilde y) & x>0,
\end{array}
\right.
\end{array}
\end{equation}
where $\bm{V}=[\bm{W};z_a]$, $\bm{F}=[\bm{f},\bm{g}]$, $\bm{V}_-(x,y)$ and $\bm{V}_+(x,y)$ are two polynomials defined on the two neighboring computational cells at time $t = 0$, respectively.
Since we just want to construct  fluxes  normal to cell interfaces, the tangential effect can be regarded as
a source term. Therefore, we rewrite the quasi 1-D GRP \eqref{Q1-GRP} as
\begin{equation}
\begin{array}{l}
\bm{W}_t+\bm{f}(\bm{W})_x=-\bm{g}(\bm{W})_y,\\[3mm]
(z_{a})_t + u (z_{a})_x - \omega u_x = \omega v_y -v (z_{a})_y,\\[3mm]
\bm{V}(x,\tilde y,0)=\left\{
\begin{array}{ll}
\bm{V}_-(x,\tilde y), \ \ \ \ & x<0, \\
\bm{V}_+(x,\tilde y), & x>0,
\end{array}
\right.
\end{array}
\label{Q2-GRP}
\end{equation}
by fixing a $y$-coordinate.  
That is, we solve the 1-D GRP at a point $(0,\tilde y)$ on the interface, by considering the transversal effect to the interface $x=0$.  The value $\bm{g}(\bm{W})_y$ and $\omega v_y -v (z_{a})_y$ at $(0,\tilde y)$  takes account of the local wave propagation.
The solution of this GRP is denoted as $\textbf{GRP}\left(\bm{V}_-(\bm{x}),\bm{V}_+(\bm{x})\right)$ and solved by a 2-D GRP solver.
This appendix introduces the 2-D GRP solver  used in the coding process just for completeness and readers' convenience. The details can be found  in \cite{qi_2017}.
We notice that the equation of $z_a$ is very close to the equation of mass fraction for the burnt gas in the basic "combustion model" \cite{ben-artzi_generalized_1989}, 
and the GRP solver for the combustion model in \cite{ben-artzi_computation_1990,ben-artzi_generalized_1989} is a heuristic  form of our 2-D GRP solver.

The GRP solver for solving \eqref{Q1-GRP}  has the following two versions, which are the acoustic version and the genuinely nonlinear version.

\subsection*{2-D acoustic case.}\label{A1}
At any point  $(0,\tilde y)$, if $\bm{V}_-(0-0,\tilde y)\approx \bm{V}_+(0+0,\tilde y)$ and $\big\|\frac{\partial \bm{V}_-}{\partial x}(0-0,\tilde y)\big\|\neq \big\|\frac{\partial \bm{V}_+}{\partial x}(0-0,\tilde y)\big\|$, we view it as an acoustic case. Denote $\bm{V}_*: = \bm{V}_-(0-0,\tilde y)\approx \bm{V}_+(0+0,\tilde y)$, and then linearize the governing equations \eqref{eq:5-eq} to get
\begin{equation}
\frac{\partial \bm{V}}{\partial t}+\bm{A}(\bm{V})\frac{\partial \bm{V}}{\partial x}+\bm{B}(\bm{V})\frac{\partial \bm{V}}{\partial y}=\bm{0}.
\end{equation}
We make the decomposition $\bm{A}(\bm{V}_*) =\bm{R} \bm{\Lambda} \bm{R}^{-1}$,
where  $\bm{\Lambda}=\mbox{diag}\{\lambda_i\}$, $\bm{R}$ is the (right) eigenmatrix of $\bm{A}(\bm{V}_*)$.  Then the acoustic GRP solver takes
\begin{equation}
\begin{array}{rl}
\displaystyle \left(\frac{\partial \bm{V}}{\partial t}\right)_{(0,\tilde y, 0)} =& \displaystyle  -\bm{R}\bm{\Lambda}^+ \bm{R}^{-1} \left(\frac{\partial \bm{V}_-}{\partial x}\right)_{(0-0,\tilde y)}-\bm{R} \bm{I}^+ \bm{R}^{-1} \left(\bm{B}(\bm{V}_-)\frac{\partial \bm{V}_-}{\partial y}\right)_{(0-0,\tilde y)}\\[3mm]
 &\displaystyle -\bm{R}\bm{\Lambda}^- \bm{R}^{-1} \left(\frac{\partial \bm{V}_+}{\partial x}\right)_{(0+0,\tilde y)}-\bm{R} \bm{I}^- \bm{R}^{-1} \left(\bm{B}(\bm{V}_+)\frac{\partial \bm{V}_+}{\partial y}\right)_{(0+0,\tilde y)},
\end{array}
\label{acoust}
\end{equation}
where $\bm{\Lambda}^+ =\mbox{diag}\{\max(\lambda_i,0)\}$, $\bm{\Lambda}^- =\mbox{diag}\{\min(\lambda_i,0)\}$,  $\bm{I}^+ =\frac 12 \mbox{diag}\{1+\mbox{sign}(\lambda_i)\}$, $\bm{I}^- =\frac 12 \mbox{diag}\{1-\mbox{sign}(\lambda_i)\}$.

\subsection*{2-D  nonlinear case.}


At any point $(0,\tilde y)$, if the difference $\|\bm{V}_-(0-0,\tilde y)-\bm{V}_+(0+0,\tilde y)\|$ is large, we regard it as the genuinely nonlinear case and have to solve the  2-D  GRP analytically. A key ingredient is how to understand $\bm{g}(\bm{W})_y$ and $\omega v_y -v (z_{a})_y$ at $(0,\tilde y)$. Here we construct the \textit{2-D GRP solver} by two steps.
\vspace{0.2cm}

(i) We solve the local planar 1-D Riemann problem
\begin{equation}
\begin{array}{l}
\bm{w}_t+\bm{f}(\bm{w})_x=\bm{0},\ \ \ \ t>0, \\[3mm]
(z_{a})_t + u (z_{a})_x - \omega u_x = 0,\\[3mm]
\mathbf{w}(x,\tilde y,0) =\left\{
\begin{array}{ll}
\bm{V}_-(0-0,\tilde y), \ \ \ &x<0,\\
\bm{V}_+(0+0,\tilde y), & x>0,
\end{array}
\right.
\end{array}
\label{q1d}
\end{equation}
where $\mathbf{w}=[\bm{w};z_a]$, to obtain the local Riemann solution $\bm{V}_* =\mathbf{w}(0,\tilde y,0+0)$.  Just as in the acoustic case, we decompose $\bm{A}(\bm{V}_*) =\bm{R}\bm{\Lambda}\bm{R}^{-1}$.  Then we set
\begin{equation}
\bm{h}(x,y)=
\begin{bmatrix}
-\overline{ \bm{g}(\bm{W})_y}\\
\overline{ \omega v_y -v (z_{a})_y}
\end{bmatrix}
 = \left\{
\begin{array}{ll}
-\bm{R} \bm{I}^+ \bm{R}^{-1} \left(\bm{B}(\bm{V}_-)\frac{\partial \bm{V}_-}{\partial y}\right)_{(0-0,\tilde y)}, \ \  \  &x<0,\\[3mm]
-\bm{R} \bm{I}^- \bm{R}^{-1} \left(\bm{B}(\bm{V}_+)\frac{\partial \bm{V}_+}{\partial y}\right)_{(0+0,\tilde y)}, \ \  \  &x>0,
\end{array}
\right.
\end{equation}
where $\bm{I}^\pm$ are defined the same as in \eqref{acoust}.
\vspace{0.2cm}

(ii)  We solve the quasi 1-D GRP
\begin{equation}\label{eq:GRP-source}
\begin{array}{l}
\bm{W}_t+\bm{f}(\bm{W})_x=-\overline{ \bm{g}(\bm{W})_y},\ \ \ \ t>0, \\[3mm]
(z_{a})_t + u (z_{a})_x - \omega u_x = \overline{ \omega v_y -v (z_{a})_y},\\[3mm]
\mathbf{w}(x,y, 0) =\left\{
\begin{array}{ll}
\bm{V}_-(x,y), \ \ \ &x<0,\\
\bm{V}_+(x,y), & x>0,
\end{array}
\right.
\end{array}
\end{equation}
to obtain $\left(\frac{\partial \bm{V}}{\partial t}\right)_*=\frac{\partial \bm{V}}{\partial t}(0,\tilde y, 0+0)$.

This is done by solving the 1-D GRP for homogeneous equations
\begin{equation}\label{eq:GRP-homo}
\begin{array}{l}
\bm{w}_t+\bm{f}(\bm{w})_x=\bm{0},\ \ \ \ t>0, \\[3mm]
(z_{a})_t + u (z_{a})_x - \omega u_x = 0,\\[3mm]
\mathbf{w}(x,\tilde y, 0) =\left\{
\begin{array}{ll}
\bm{V}_-(x,\tilde y), \ \ \ &x<0,\\
\bm{V}_+(x,\tilde y), & x>0,
\end{array}
\right.
\end{array}
\end{equation}
For the augmented Euler equations, $\left(\frac{\partial\mathbf{w}}{\partial t}\right)_*=\frac{\partial\mathbf{w}}{\partial t}(0,\tilde y, 0+0)$ is obtained by solving a pair of algebraic equations essentially,
 \begin{equation}
 \begin{array}{ll}
\displaystyle a_L\left(\frac{\partial u}{\partial t}\right)_* + b_L \left(\frac{\partial p}{\partial t}\right)_*=d_L,\\[3mm]
\displaystyle  a_R\left(\frac{\partial u}{\partial t}\right)_* + b_R\left(\frac{\partial p}{\partial t}\right)_*=d_R.
 \end{array}
 \end{equation}
At last, we have 
 \begin{equation*}
\left(\frac{\partial\bm{V}}{\partial t}\right)_*=\left(\frac{\partial\mathbf{w}}{\partial t}\right)_*+\bm{h}(x,\tilde y).
 \end{equation*}
Here, if we ignore $\bm{h}(x,\tilde y)$, the solver is called \textit{1-D GRP solver}.
Solvers for the GRP \eqref{eq:GRP-source} with a general source term is presented in \cite{Li-2}. When specified to \eqref{Q2-GRP}, construction of the solver can be found in \cite{qi_2017}.

\section*{Acknowledgement} This research is supported by the Natural Science Foundation of China (11771054, 91852207,12072042), National Key Project (GJXM 92579) and Foundation of LCP. We appreciate Professor Matania  Ben-Artzi for his  many kind comments.  
\bibliographystyle{mybst}
\bibliography{ES-bib}

\end{document}